\documentclass[prl,twocolumn,superscriptaddress]{revtex4-1}
\newcommand{\Eq}[1]{Eq.~(\ref{eq:#1})}

\newcommand{\Fig}[1]{Fig.~\ref{fig:#1}}
\newcommand{\Figs}[1]{Figs.~\ref{fig:#1}}

\newcommand{\Ref}[1]{Ref.~\cite{#1}}
\newcommand{\expt}[1]{\left<#1\right>}

\newcommand{\n}{\mathbf{n}}
\newcommand{\F}{\mathbf{F}}
\renewcommand{\v}{\mathbf{v}}
\newcommand{\rr}{\mathbf{r}}
\newcommand{\rms}{\mathrm{rms}}
\newcommand{\rrijC}{\rr_{i,j}^C}
\newcommand{\gdot}{\dot\gamma}

\newcommand{\nij}{\hat\n_{ij}}
\newcommand{\omegabody}{\widetilde\omega}
\newcommand{\skippa}[1]{}
\usepackage{graphicx}
\usepackage{amsmath}
\usepackage{amssymb}
\usepackage{color}

\begin{document}
\title{Translational and rotational velocities in shear-driven jamming of ellipsoidal particles}
\author{Yann-Edwin Keta}
\affiliation{Department of Physics, Ume\aa\ University, 901 87 Ume\aa, Sweden}
\affiliation{D\'epartement de Physique, \'Ecole Normale Sup\'erieure de Lyon, 69364 Lyon Cedex
  07, France}
\affiliation{D\'epartement de Physique, Universit\'{e} Claude Bernard Lyon 1, 69622
  Villeurbanne Cedex, France
}

\author{Peter Olsson}
\affiliation{Department of Physics, Ume\aa\ University, 901 87 Ume\aa, Sweden}

\date{\today}

\begin{abstract}
  We study shear-driven jamming of ellipsoidal particles at zero temperature with a focus
  on the microscopic dynamics. We find that a change from spherical particles to
  ellipsoids with aspect ratio $\alpha=1.02$ gives dramatic changes of the microscopic
  dynamics with much lower translational velocities and a new role for the
  rotations. Whereas the velocity difference at contacts---and thereby the
  dissipation---in collections of spheres is dominated by the translational velocities and
  reduced by the rotations, the same quantity is in collections of ellipsoids instead
  totally dominated by the rotational velocities. By also examining the effect of
    different aspect ratios we find that the examined quantities show either a peak or a
    change in slope at $\alpha\approx1.2$, thus giving evidence for a crossover between
    different regions of low and high aspect ratio.
\end{abstract}


\maketitle

\paragraph{Background}
Dense collections of circular disks in two dimensions (2D) and spheres in 3D with contact
interaction at zero temperature have been studied extensively during the last decades with
the aim to understand the jamming transition. Since ordinary molecular dynamics that
automatically explores phase space doesn't work at zero temperature one has mainly used
two other computational methods. The first is what we call isotropic jamming, which is the
generation of particle packings with energy-minimization methods in various
ways\cite{Donev_TS:packing, Berthier_Witten:PRE2009, Vagberg_OT:protocol, Ozawa:2017,
  OHern_Langer_Liu_Nagel:2002, OHern_Silbert_Liu_Nagel:2003, Vagberg_OT:protocol} and the
second is to model a shear flow by steadily shearing the system, either quasistatically
\cite{Vagberg_VMOT:jam-fss, Heussinger_Barrat:2009,
  Heussinger_Chaudhuri_Barrat-Softmatter} or at different constant shear rates
\cite{Olsson_Teitel:jamming,Hatano:2008, Hatano:2011, Otsuki_Hayakawa:2009b,
  Hatano:2009, Hatano:2010, Tighe_WRvSvH}.

Since real particles are seldom perfectly spherical an interesting generalization of this
model with obvious experimental relevance is the change from spherical to aspherical
particles. It has already been argued that the spherical limit is singular
\cite{Marschall_Keta_Olsson_Teitel:2019, Marschall_Teitel:sph-cyl-2018,
  Marschall_Teitel:sph-cyl-2019, Marschall_Teitel:sph-cyl-2020} and the purpose of the
present Letter is to explore further consequences of this change to aspherical particles.

A key result from the study of jamming of spherical particles is that jamming occurs when
the average number of contacts per particle is just enough to stabilize the degrees of
freedom, according to the isostatic condition, $z=2d_f$. Since rotations become relevant
for aspherical particles we find $d_f=6$ for general ellipsoids and $d_f=5$ for ellipsoids
with a symmetry axis (spheroids), which is the kind of particles studied in the present
work. If the isostatic conjecture were always valid one would expect a jump in $z$ already
for a slight deviation from the spherical limit. It has however long been known that there
is no jump in $z$ \cite{Donev-Elli:2004} and that the systems are thus hypostatic at
jamming, i.e.\ have a smaller number of contacts than number of degrees of freedom.

To see how a hypostatic packing can be jammed it is helpful to consider a contact between
two ellipses that barely touch at their respective waists. This contact will then not only
hinder the motion of the ellipses toward each other but will also hinder their respective
rotations, thus affecting two different degrees of freedom for each particle. Whereas the
interaction energy will be quadratic in the position coordinate it will instead be
quartic---to the fourth power---in the angle coordinates. To see this, consider an
ellipse, $x^2+a^2y^2=r_0^2$ with $a<1$ and thus elongated along the $y$ axis, as
illustrated in the Supplemental Material (SM). The distance from the origin to the
perimeter of the ellipse is then $r^2=r_0^2+(1-a^2)y^2$, which for small $y$, such that
the angle is $\theta=y/r_0$, becomes $r(\theta)/r_0-1\sim \theta^2$. With two ellipses
barely touching at their respective waists, $\theta=0$, the effect of a rotation is an
overlap $\sim\theta^2$, and the fourth power follows since the energy is in turn quadratic
in the overlap. For a jammed collection of particles this mechanism leads to a set of
quartic vibrational modes \cite{Donev-Elli:2007, Mailman:ellipse-jamming,
  Schreck:ellipsoidal}, beside the ordinary modes for which the energy is quadratic in the
displacements. (For a particle with several contacts, not at their waists, $r(\theta)$ for
each contact also has a term linear in the rotation $\theta$. In force-balanced jammed
packings we expect these linear terms to cancel each other out, such that only the
$\theta^2$-dependence remains.)  From studies of static packings it has been found that
the number of quartic modes exactly matches the deviation in contact number from the
isostatic value \cite{Mailman:ellipse-jamming}.

In the present Letter we do shearing simulations of both spherical and ellipsoidal
particles to examine how the asphericity affects the microscopic dynamics.  We do find
dramatic effects. Focusing first on aspect ratio $\alpha=1.02$ we find that the
ellipsoidal particles have considerably lower translational velocities and that their
rotational motion gets a different role. Since the quartic modes found in static packings
are primarily rotational in character \cite{Mailman:ellipse-jamming, Schreck:ellipsoidal}
we believe that these effects are consequences of the quartic modes on the shear-driven
dynamics.  Our work thus opens up a new window for the study of shear-driven elongated
particles, which until now have mostly been analyzed with focus on possible ordering of
the particles \cite{B-Stannarius:2012, Marschall_Keta_Olsson_Teitel:2019,
  Marschall_Teitel:sph-cyl-2020} and macroscopic quantities \cite{Nagy:2017,
  Trulsson:2018, Marschall_Teitel:sph-cyl-2019}. When doing simulations also with a set of
larger aspect ratios we find several quantities to have features at $\alpha\approx1.2$,
which suggests the existence of different regions of high and low aspect ratio.

\paragraph{Model}

Our system is a bidisperse collection of particles with nominal diameters $d_b=1.4 d_s$,
in equal proportion. The particles are prolate spheroids, i.e.\ ellipsoids with two equal
minor axes with $d^{(1)}>d^{(2)}=d^{(3)}$. The aspect ratio is
$\alpha=d^{(1)}/d^{(2)}>1$. To make the particle volume independent of $\alpha$ we take
the semi-axes to be given by $d^{(1)} = \alpha^{2/3} d$ and
$d^{(2)} = d^{(3)} = \alpha^{-1/3} d$, where $d=d_b$, $d_s$ for big and small particles.
The method used to check for particle overlaps is described in
\Ref{Marschall_Keta_Olsson_Teitel:2019}.  For overlapping ellipsoids we define a scale
factor $\delta<1$ such that the ellipsoids are just barely touching when they are rescaled
by $\delta$, keeping the center of mass position fixed. The elastic force between
overlapping particles is then given by
$\F^\mathrm{el}_{ij} = k_e \delta(1-\delta) \nij/ [(\rr_i-\rr_j)\cdot\nij]$ where $\nij$
is a unit vector pointing inwards to particle $i$ at the point of contact with particle
$j$.

\paragraph{Simulations}

We take a purely collisional dynamics where dissipation takes place at the particle
contacts. With $\rrijC\equiv\rr_{j,i}^C$ the position of the point of contact of particle
$i$ with particle $j$---a point on the rescaled ellipsoids---the velocity of the
surface of particle $i$ at this point is
\begin{equation}
  \v_i(\rrijC) =  \v_i + \v_i^R(\rrijC),
  \label{eq:vi.at.rijC}
\end{equation}
where $\v_i$ is the center of mass translational motion and
$\v_i^R(\rrijC)= \boldsymbol{\omega}_i\times (\rrijC - \rr_i)$ is the velocity at the
point of contact due to $\boldsymbol\omega_i$, the rotational velocity of particle $i$.  The
velocity difference at that point is $\v^C_{ij} = \v_i(\rrijC) - \v_j(\rrijC)$ and the
dissipative force is given by $\F^\mathrm{dis}_{ij} = -k_d\v^C_{ij}$.  The equations of
motion are
\begin{eqnarray}
  m_i\dot\v_i & = & \sum_j [\F^\mathrm{el}_{ij} + \F^\mathrm{dis}_{ij}], \label{eq:mvdot}\\
  \mathbf{I}_i\cdot\dot{\boldsymbol\omega}_i & = & \sum_j (\rrijC - \rr_i)
  \times [\F^\mathrm{el}_{ij} + \F^\mathrm{dis}_{ij}],\label{eq:Iomegadot}
\end{eqnarray}
where $\mathbf{I}_i$ is the moment of inertia tensor. This model is one of the simplest
with a reasonable dynamics but it is unusual in that it has no coupling between the
tangential dissipation and the strength of the elastic force. As shown in the SM the
introduction of such a coupling however only gives minor changes, at the densities of
interest, simply because the dissipating forces are there anyway typically much smaller
than the elastic ones.  We take $k_e=1$ and $k_d=1/2$ and simulate with $N=1024$
particles. The density (filling factor) is $\phi= N(d_s^3 + d_b^3)\pi/(12L^3)$. We take
the unit of length to be equal to $d_s$, the unit of energy to be $k_e$ and the unit of
mass to be $m_s$, and let the particle mass be proportional to the particle volume. The
unit of time is $t_0=d_s\sqrt{m_s/k_e}$.

\paragraph{Asphericity and shear rate dependence}

Before turning to the microscopic behavior a few words on the shear rate dependence of
slightly aspherical particles is in order. At high shear rates such particles behave
mostly as spherical particles but as the shear rate is lowered the behavior crosses over
to a low shear rate behavior that is controlled by the aspericity \cite{Ikeda:2020}. The
reason for this is that typical particle overlaps at high shear rates may be bigger than
the typical deviation of the particle surface from the spherical whereas they become much
smaller at low shear rates. The particles then have time to rotate to pack more densely
and the behavior is then controlled by the asphericity. Since the jamming density is
expected to increase linearly with the distance from the spherical point,
$\phi_J(\alpha)-\phi_J(1)\sim|\alpha-1|$, as found from isotropic jamming
\cite{Donev-Elli:2004}, one expects the main effect of the asphericity close to jamming to
be to shift the curves to (somewhat) higher $\phi$
\cite{Marschall_Teitel:sph-cyl-2019}. Fig.~S3 in the SM illustrates this crossover for the
shear viscosity.

\begin{figure}
  % plo vomega-phi.plo
  \includegraphics[width=4.2cm, bb=50 324 360 660]{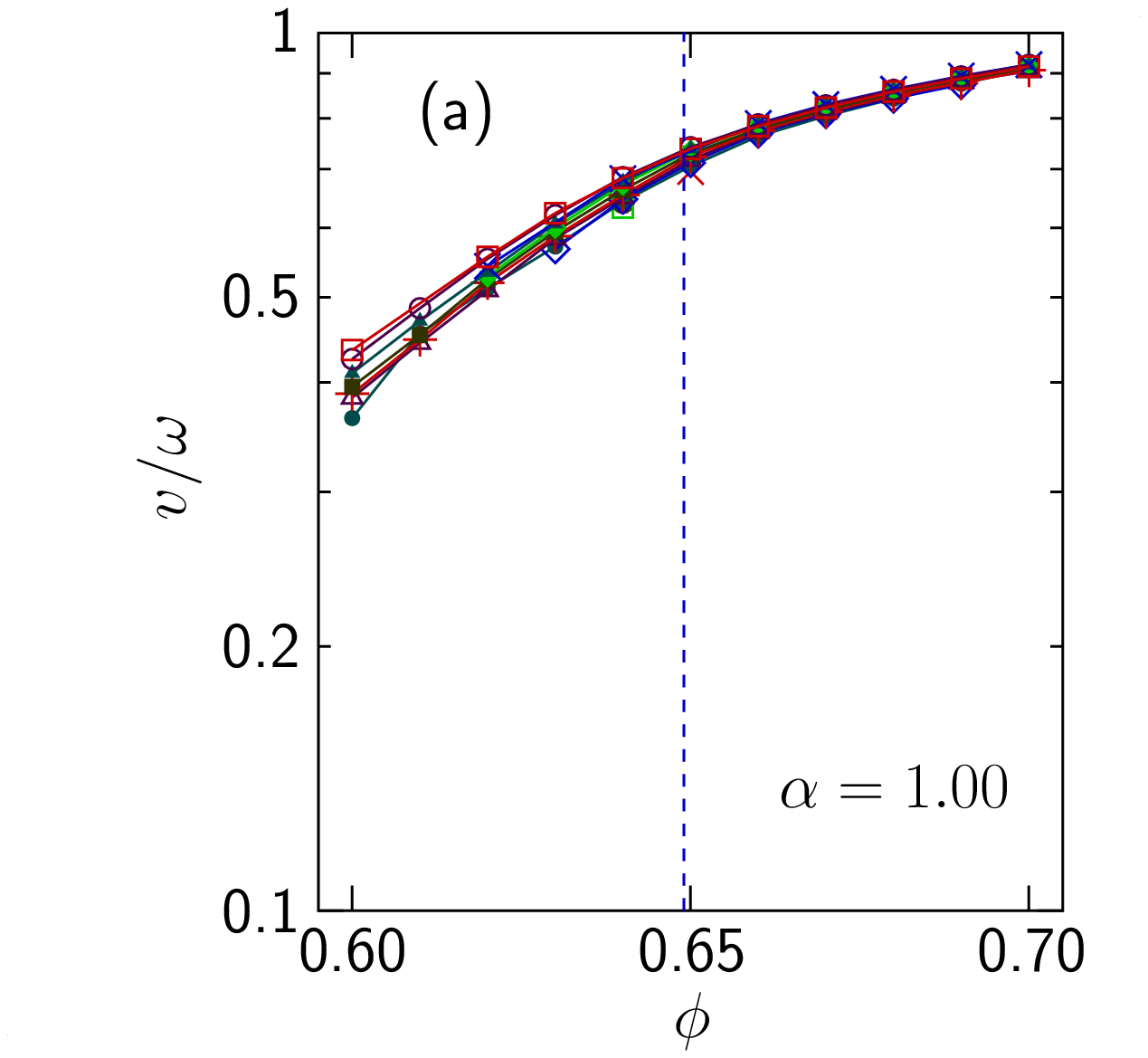}
  \includegraphics[width=4.2cm, bb=50 324 360 660]{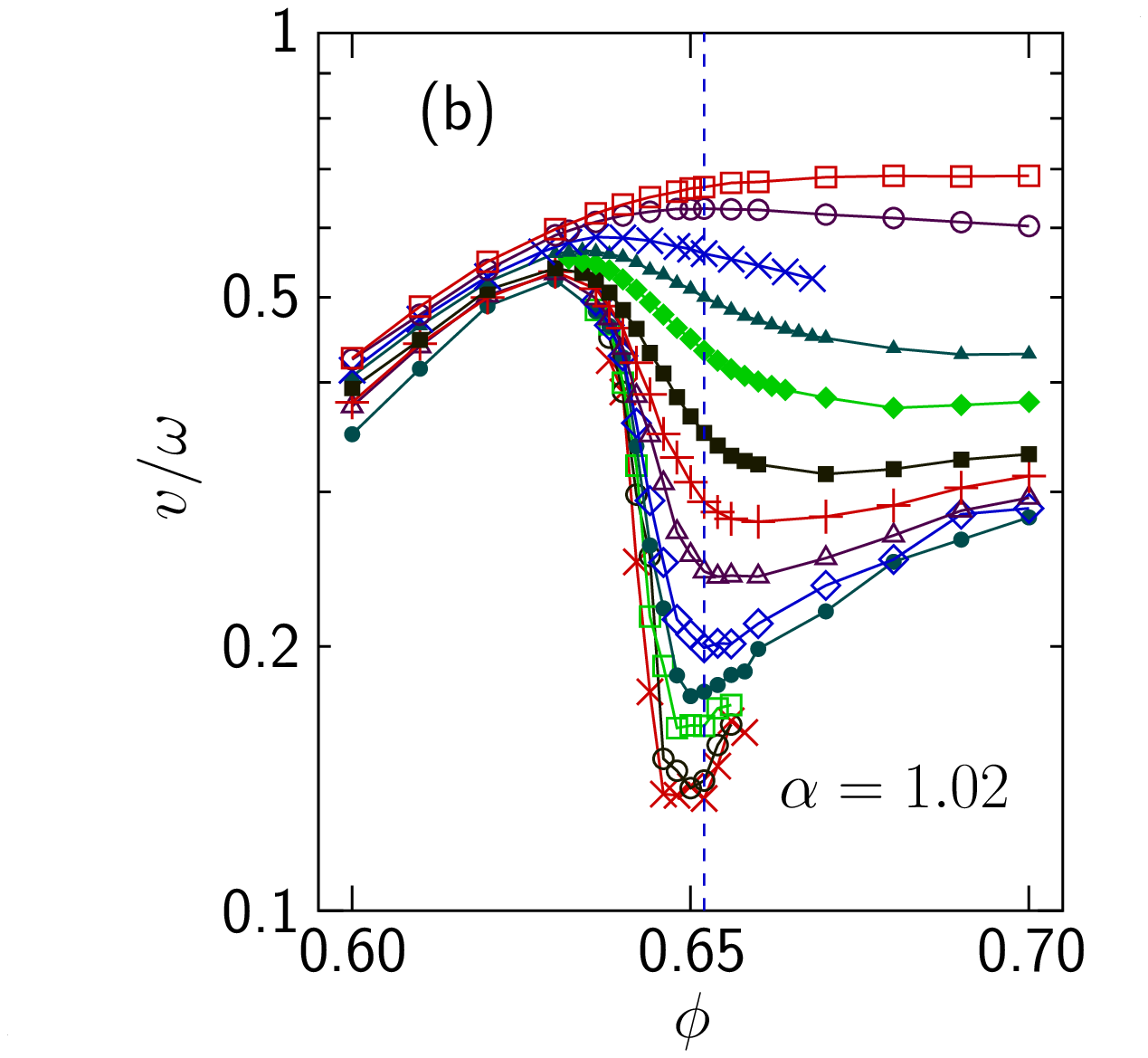}
  % plo vomega-gdot-6520-102.plo
  \includegraphics[width=4.2cm, bb=50 324 360 660]{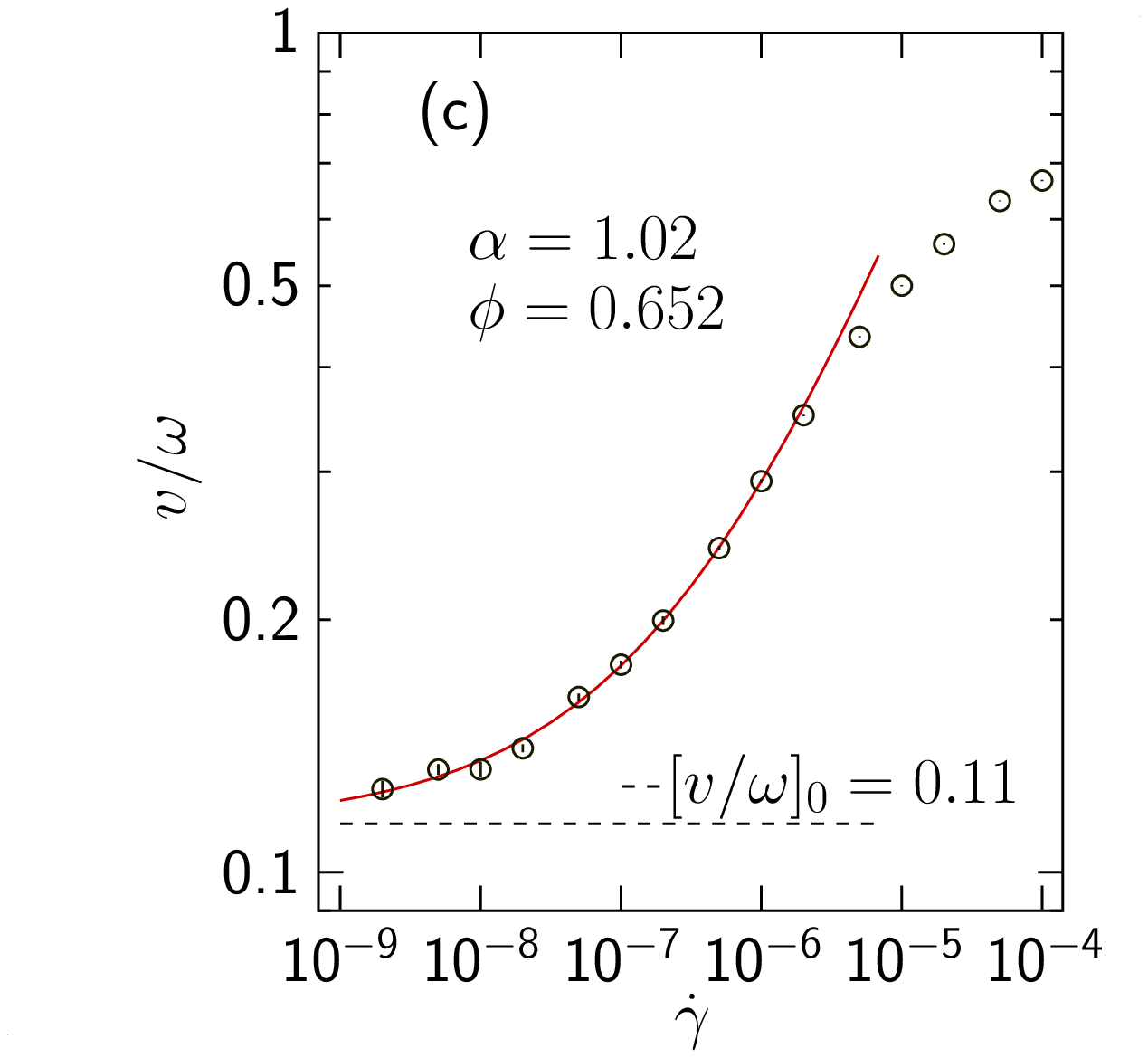}
  \includegraphics[width=4.2cm, bb=50 324 360 660]{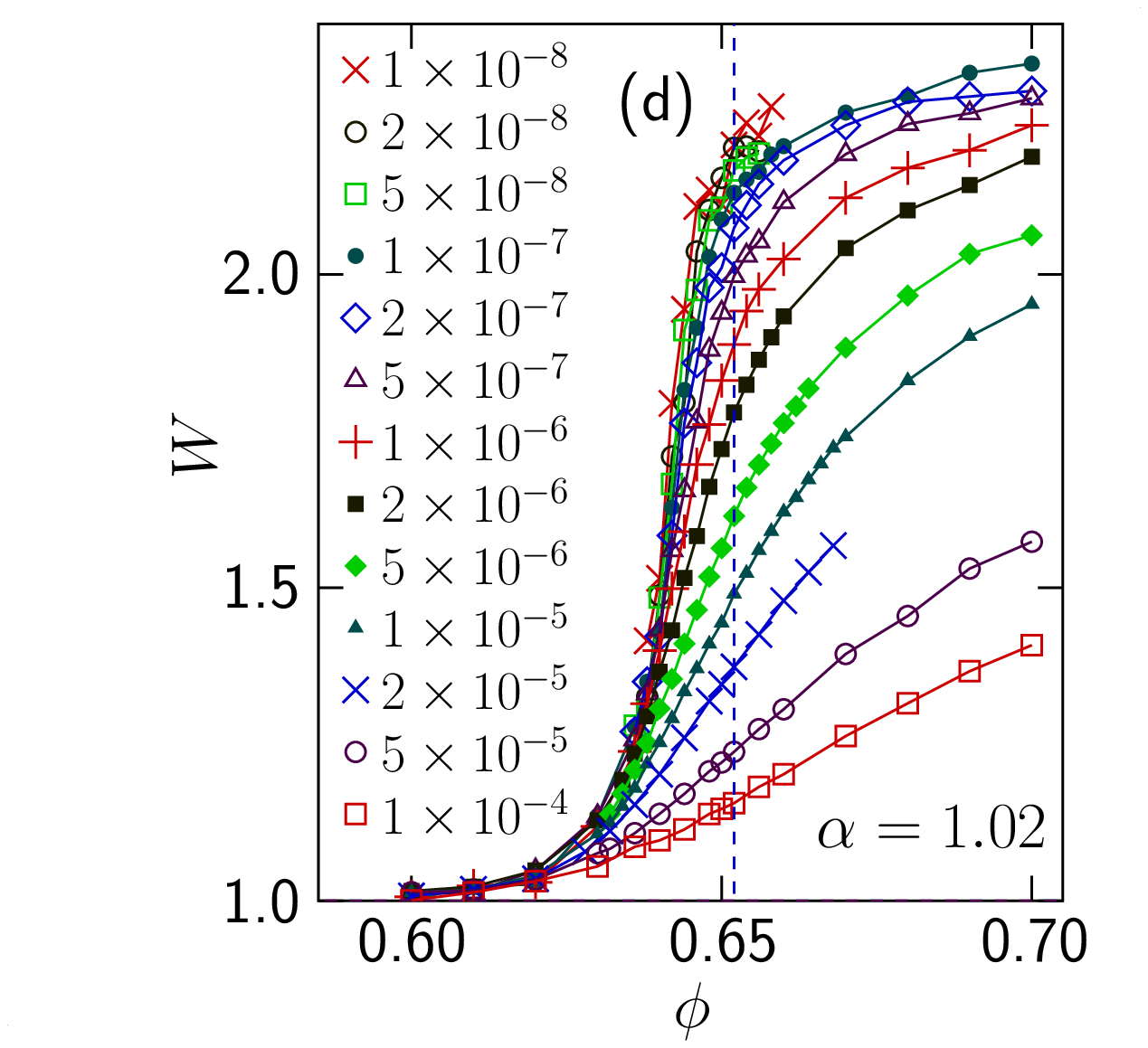}
  \caption{Measures of translational and rotational particle velocities for spheres,
    $\alpha=1.00$ and ellipsoids with $\alpha=1.02$. Panels (a) and (b) show $v/\omega$ vs
    $\phi$ for spheres and ellipsoids. (Legends for the different $\gdot$ are in panel
    (d).) For spheres this quantity only depends slowly on $\phi$ but for ellipsoids it
    decreases in a dramatic way, and even more so for the lower shear rates. The vertical
    dashed lines are the respective $\phi_J$.  Panel (c) shows an extrapolation of
    $v/\omega$ at $\phi_J(1.02)$ to the $\gdot\to0$ limit, giving
    $[v/\omega]_0\approx 0.11$.  Panel (d) shows $W$ which is the ratio of the
    root-mean-square rotational velocities along the minor axes and the major axis, as
    described in the main text. The figure shows that the rotations are predominantly
    around the minor axes which are the ones that can help the particles fit in with their
    neighbors.}
  \label{fig:R}
\end{figure}

\paragraph{Translational and rotational velocities}

A study of the microscopic dynamics however shows that the onset of particle asphericity
also gives other effects than shifting the jamming transition.  A simple and direct way to
investigate the dynamics is to determine the ratio of the non-affine translational
velocity and the rotational velocity, $v/\omega$. $v$ and $\omega$ are here the
root-mean-square non-affine translational velocity
$\v_i^\mathrm{na}=\v_i-\gdot y_i\hat{x}$ and the root-mean-square rotational velocity
$\boldsymbol\omega_i$, i.e.\ $v=\sqrt{\expt{(\v_i^\mathrm{na})^2}}$, and
$\omega=\sqrt{\expt{\boldsymbol\omega_i^2}}$.  \Figs{R}(a) and (b) show the ratio
$v/\omega$ vs $\phi$ for spherical and ellipsoidal particles at several different shear
rates. For spherical particles, panel (a), this ratio is to a good approximation
independent of $\gdot$ and takes values from 0.4 through 0.9 in the density window of the
figure. For ellipsoidal particles, panel (b), the behavior is similar to that of the
spheres at low densities but becomes very different at higher densities, with a dramatic
shear rate dependence and a big drop. The ratio $v/\omega$ changes by a factor of about
five for our range of shear rates. As this is for an asphericity $\equiv \alpha-1$ of only
2\%, the effect is surprisingly big. As shown in the SM the decrease in this ratio is
largely an effect of a smaller $v$ but there is also a contribution from an increase in
$\omega$.

To study the dependence on $\alpha$ we compare data at the respective jamming densities,
$\phi_J(\alpha)$. For $\alpha=1.00$ we take $\phi_J(1.00)\approx0.648$
\cite{Olsson:jam-3D} and for $\alpha>1$ the $\phi_J(\alpha)$ are determined as the
densities that give similar algebraic decays of $p(\gdot)$. Details are given in the
SM. For the slightly aspherical particles this gives $\phi_J(1.02)=0.652$.  We stress that
these are approximate values since $\phi_J$ is always difficult to determine, and this is
even more the case for $\alpha$ close to the spherical limit. These (small) uncertainties
in $\phi_J(\alpha)$ are however not of any big concern since the studied quantites depend
only slowly on $\phi$.

With the pronounced decrease of $v/\omega$ with decreasing $\gdot$ in \Fig{R}(b) it
becomes interesting to try to extrapolate to the limit of vanishing shear rate. At
$\phi_J(1.02)=0.652$---which is also the density where $v/\omega$ has its minimum---we fit
to constant plus algebraic behaviour,
$v(\phi_J, \gdot)/\omega(\phi_J, \gdot) = [v/\omega]_0 + a\gdot^b$.  As shown in
\Fig{R}(c) the fit is good and suggests that the limiting value is a finite constant,
$[v/\omega]_0\approx0.11$. This quantity in also shown in \Fig{vs-alpha}(b) for general
$\alpha$.

\paragraph{Rotations around different axes}
Slowly sheared spherical particles close to jamming display erratic translational
displacements which are needed to fit into a constantly changing environment. The lower
translational velocity of the ellipsoidal particles suggests a picture where this
translational motion is largely replaced by rotations. Since only the rotations around the
minor axes can help the particles fit in with their neighbors, one expects these rotations
to be more prominent than the rotations around the major (symmetry) axis. With the
rotation vector in the body frame given by
$\widetilde{\boldsymbol\omega}= (\omegabody^{(1)}, \omegabody^{(2)}, \omegabody^{(3)})$,
where the major axis is direction ``1'', \Fig{R}(d) shows
$W\equiv[\omegabody_\rms^{(2)}+\omegabody_\rms^{(3)}]/2\omegabody_\rms^{(1)}$---from the
respective root-mean-square values---vs $\phi$ for different $\gdot$.  At low densities
this quantity is close to unity, indicating that the rotations are equally strong around
the different semi-axes, but it increases to higher values around the jamming
density. This increase is small for high $\gdot$, reaches more than a factor of two for
our lowest $\gdot$, and an extrapolation at $\phi_J$, again by fitting to constant plus
algebraic behavior, suggests the $\gdot\to0$ limit $W_0=2.25$.

\begin{figure}
  % plo grc-show.plo
  \includegraphics[bb=11 324 532 657, width=7cm]{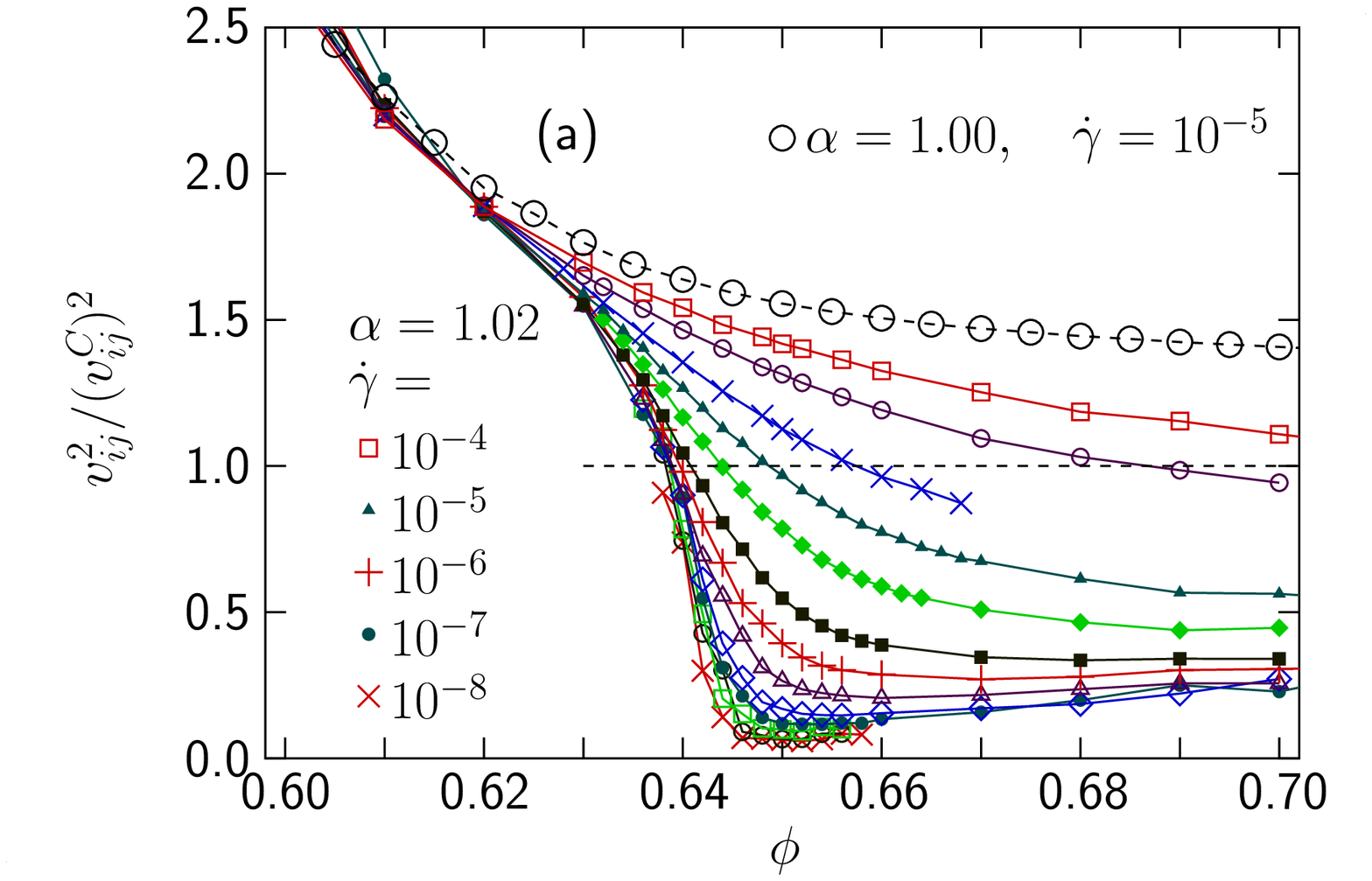}
  \includegraphics[width=4.2cm, bb=50 324 360 660]{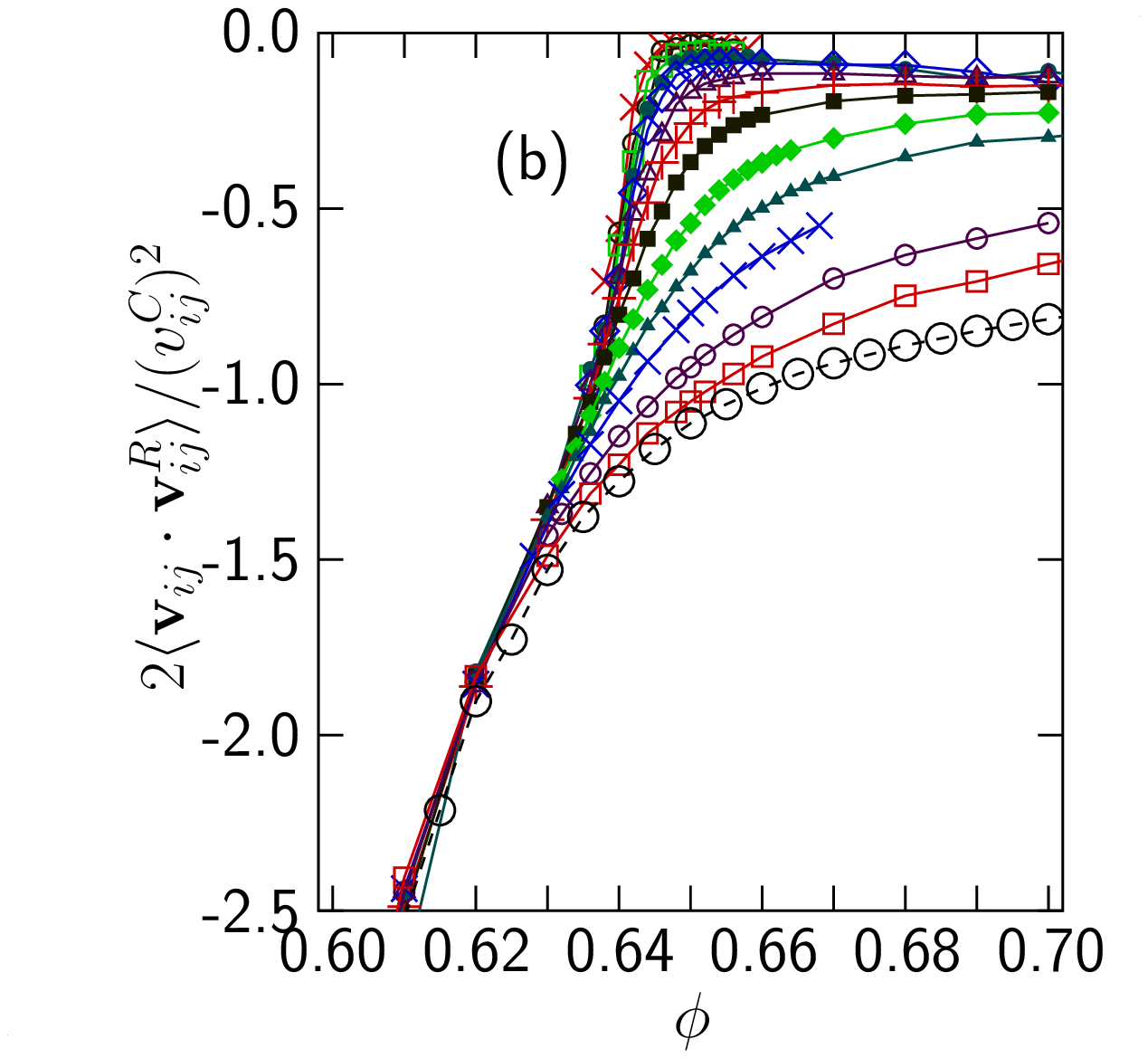}
  \includegraphics[width=4.2cm, bb=50 324 360 660]{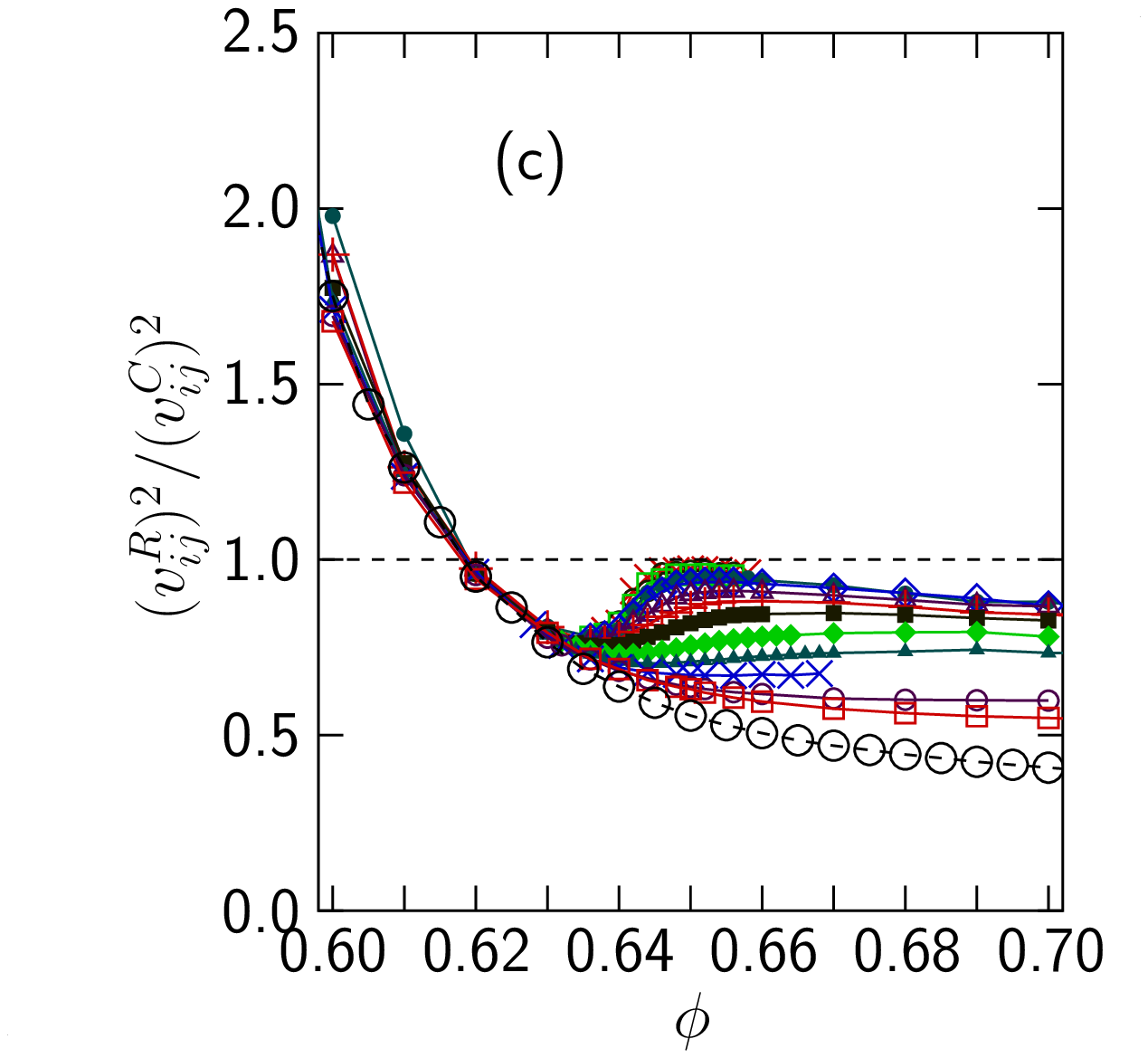}
  \caption{Contributions of the different terms of \Eq{vCij2} to $(v^C_{ij})^2$---and thus
    to the dissipation---for both spheres (open circles, $\gdot=10^{-5}$, no significant
    $\gdot$-dependence) and ellipsoids for many different shear rates,
    $\gdot=1\times10^{-8}$, $2\times10^{-8}$, $5\times10^{-8}$,\ldots, $10^{-4}$. The
    figure shows the difference between spheres and ellipsoids in two ways: (i) For
    spheres the velocity difference is mainly due to the translational velocity difference
    and the effect of the rotations---panels (b) and (c)---is to reduce the velocity
    difference at contacts. (ii) For ellipsoids at densities around $\phi_J(1.02)=0.652$
    and in the limit of small shear rates (crosses for $\gdot=10^{-8}$) the velocity
    difference is mainly due to the rotational velocity. The translational
    velocity---panels (a) and (b)---gives a very small contribution.}
  \label{fig:vijRT}
\end{figure}

\paragraph{Velocity difference at particle contacts}

Whereas the quantities discussed above are single-particle quantities, we now turn to the
velocity difference at particle contacts, and, more specifically, the role of rotations
and translations for these velocity differences. Since the velocity difference is related
to the dissipation, and thereby to the shear viscosity, this is a quantity with a more
direct physical relevance than the single particle velocities, and it is also the quantity
that most clearly shows the difference between spherical and aspherical particles.
\skippa{The relation between velocity difference and shear stress follows from the need
  for the supplied power $P_\mathrm{in} = V\sigma\gdot$ to on average balance the
  dissipated power, $P_\mathrm{diss} = k_d \sum_{ij} (\v^C_{ij})^2$, where the sum is over
  all particles $i$ and $j$ in contact \cite{Ono_Tewari_Langer_Liu}.}

For the analysis we use the notation in and after \Eq{vi.at.rijC} and write the contact
velocity difference as the sum of translational and rotational parts,
$\v^C_{ij} = \v_{ij} + \v^R_{ij}$. The contact velocity difference squared then becomes
\begin{equation}
  \label{eq:vCij2}
  (v^C_{ij})^2 = v^2_{ij} + 2 \expt{\v_{ij} \cdot \v^R_{ij}} + (v^R_{ij})^2.
\end{equation}
The relative contributions of these three terms are shown in \Fig{vijRT} both for spheres
(where the behavior is independent of $\gdot$) and for ellipsoids at several different
$\gdot$. We then compare the behavior of ellipsoids for the lowest $\gdot$ with the
behavior of spheres. For spheres $v^C_{ij}$ is dominated by the translational velocities,
$\v_{ij}$, and reduced by the rotations, $\v^R_{ij}$, which is seen by panels (b) and (c)
together giving a negative contribution. The relative contributions of the three terms in
\Eq{vCij2} at $\phi=0.648$ are $1.57$, $-1.13$, and $0.56$.

For the ellipsoids the main contribution is from the rotations, $\v^R_{ij}$, and the
contribution from $\v_{ij}$ is very small. This is seen by the crosses, which are data for
$\alpha=1.02$ and $\gdot=10^{-8}$, in the three panels at $\phi_J$ being close to zero and
unity, respectively. After extrapolating the three terms of \Eq{vCij2} at
$\phi_J(1.02)=0.652$ to $\gdot\to0$ their relative contributions are found to be $0.04$,
$-0.02$, and $0.98$, which shows that it is the rotations that totally dominate the
velocity difference.  The translations only contribute 4\% of the dissipated power
whereas the rotations, when including the negative mixed term, contribute the remaining 96\%.

\skippa{The small contribution to the velocity difference for ellipsoid particles from
  $v_{ij}$ compared to $v^R_{ij}$ is another consequence of the low $v/\omega$, shown in
  \Fig{R}(b). That the lower translational velocity makes $v^2_{ij}$ small is no big
  surprise, but it is interesting that this is compensated for by $\v^R_{ij}$ in a way
  that leaves the total dissipation unchanged.}

\paragraph{Different aspect ratios}

We now turn to the dependence on the aspect ratio, $\alpha$, by comparing behaviors of
several quantities at the different $\phi_J(\alpha)$, extraploated (as shown in the SM) to
the $\gdot\to0$ limit. \Fig{vs-alpha}(a) shows that $W_0(\alpha)$---the relative rotation
around the minor axes---first increases with increasing $\alpha$, reaches a maximum at
$\alpha\approx1.2$, then starts decreasing and is close to unity at $\alpha=2.5$.  It is
interesting to note that this peak in $W_0(\alpha)$ not coincides with the maximum jamming
density, which is at $\alpha\approx1.5$ both in determinations from isotropically jammed
packings \cite{Donev-Elli:2004, Donev-Elli:2007} and in the present study. In panel (b)
$[v/\omega]_0$ shows a change from one linear region to another at $\alpha\approx1.2$. In
panel (c) which shows the relative contribution of the different terms in
\Eq{vCij2}---excluding $\alpha=1$ which is altogether different---the same kind of change
is also present, though not that sharp. That figure shows that the contribution to the
dissipation from the rotations which is very big at $\alpha=1.02$, decreases monotonously
with increasing $\alpha$. Panel (d) shows that the rotational velocity around the $z$ axis
(of the lab frame) associated with the shearing, which is $-[\omega_z/\gdot]_0=1/2$ for
spheres, first stays almost constant but then starts decreasing for $\alpha>1.2$.

\begin{figure}
  % plo v.omega-gdot.plo
  \includegraphics[width=4.2cm, bb=50 324 360 660]{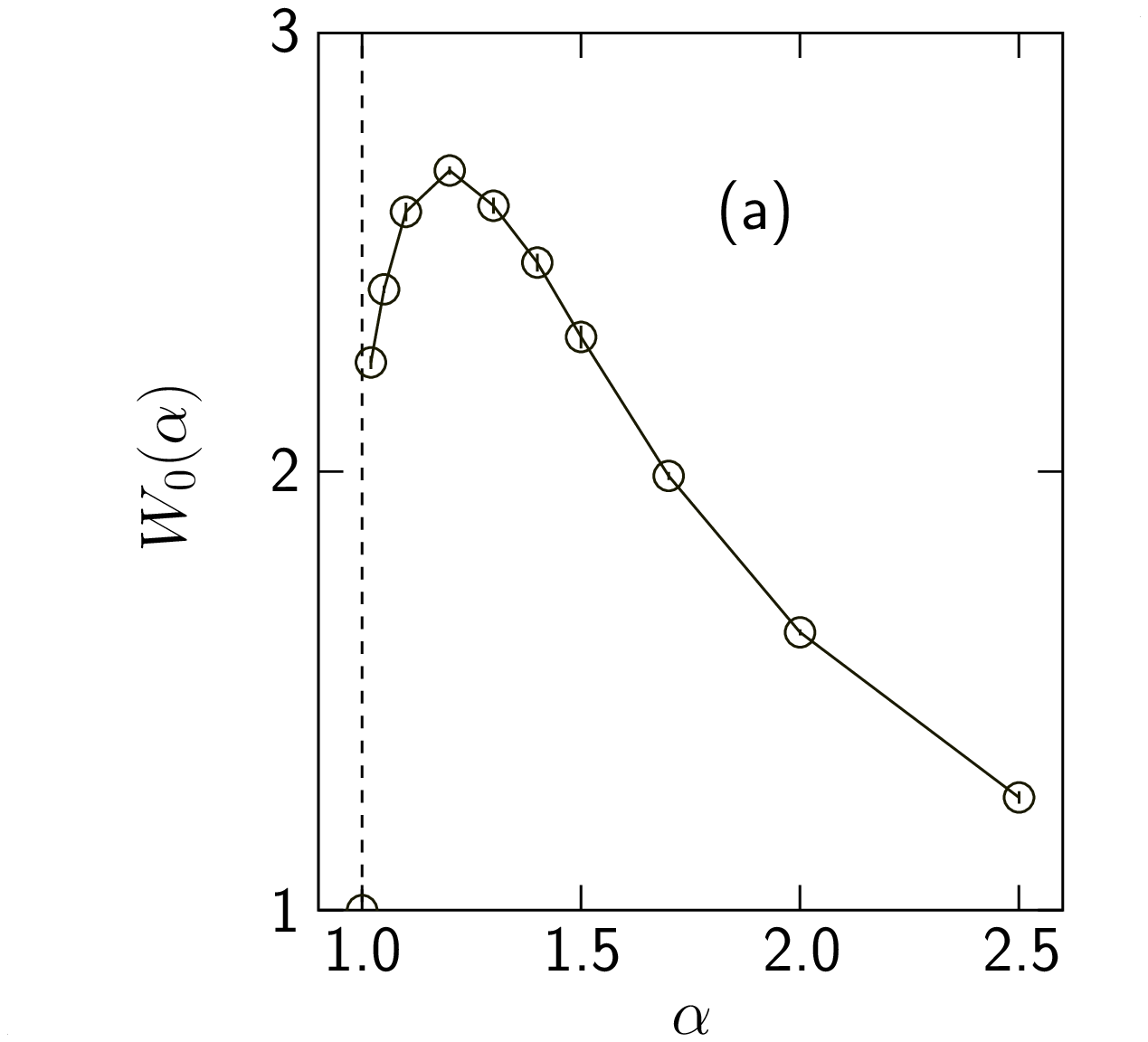}
  \includegraphics[width=4.2cm, bb=50 324 360 660]{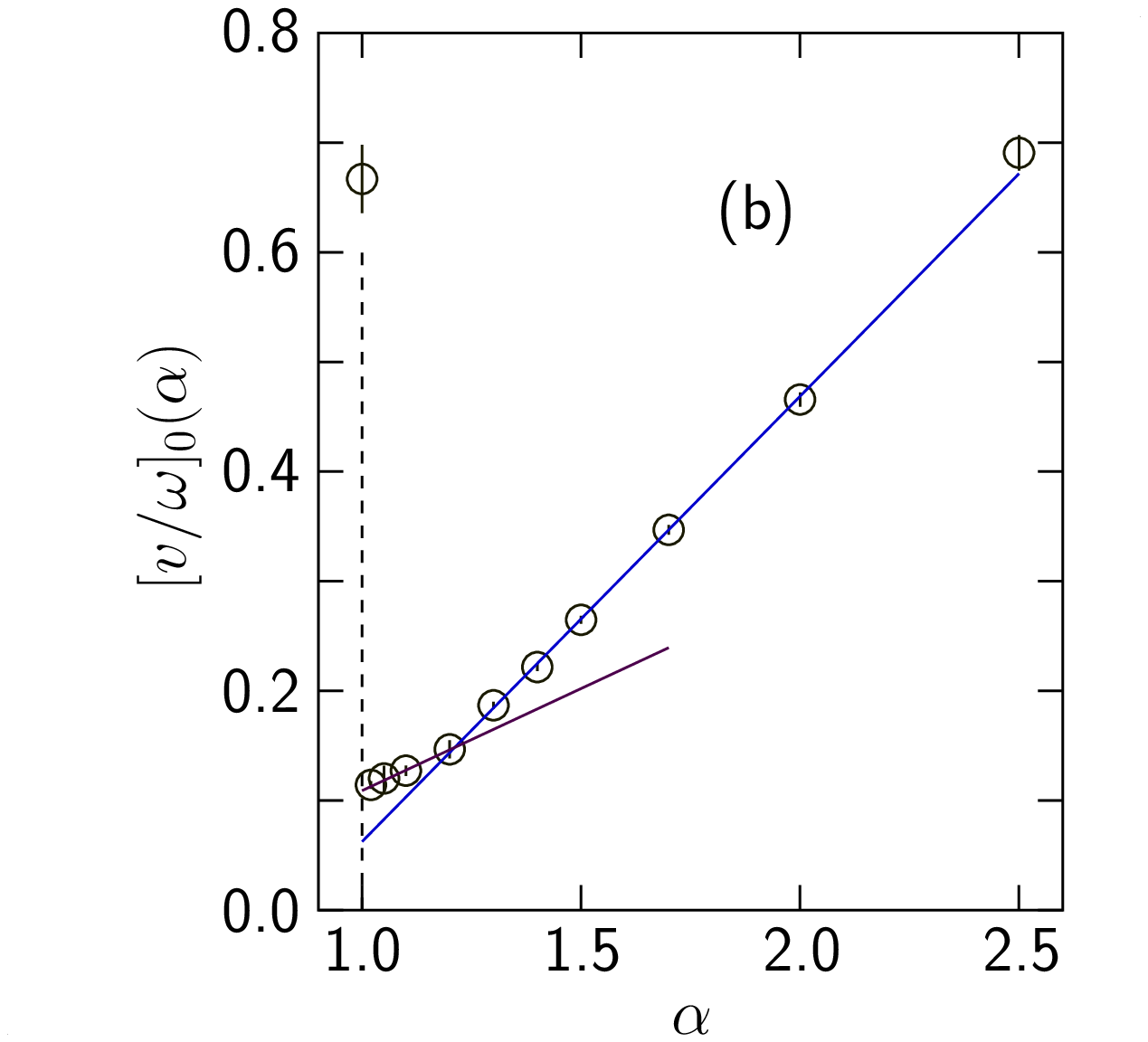}
  \includegraphics[width=4.2cm, bb=50 324 360 660]{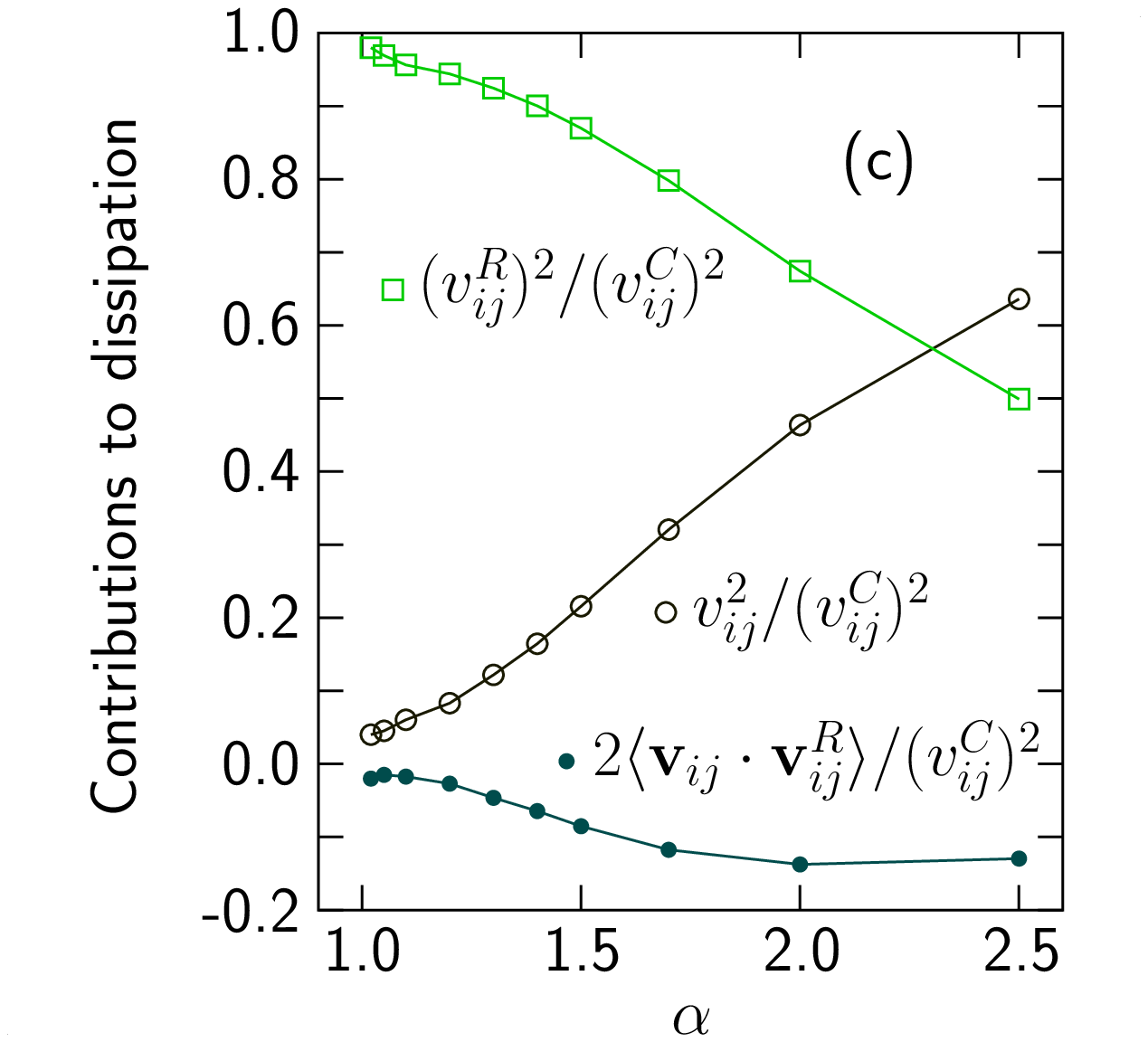}
  \includegraphics[width=4.2cm, bb=50 324 360 660]{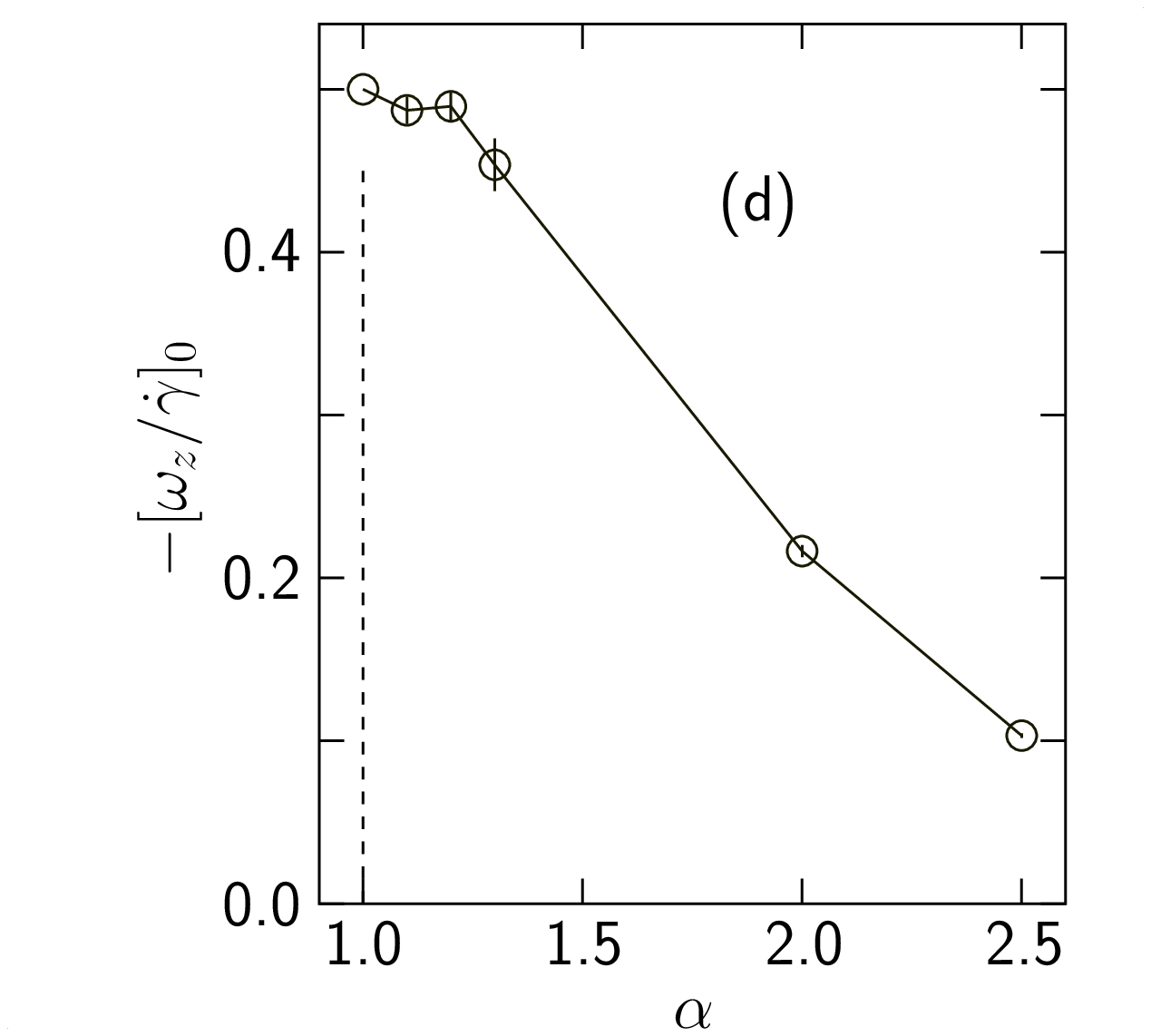}
  \caption{Dependency on the aspect ratio, $\alpha$. All quantities are obtained after
    extrapolating to the $\gdot\to0$ limit at the respective $\phi_J(\alpha)$. Panels (a)
    and (b) are $W_0$ and $[v/\omega]_0$ from \Fig{R}. Panel (c) relates to \Fig{vijRT}
    and shows the relative contribution to the dissipation of the three terms in
    \Eq{vCij2} vs $\alpha$. (The points for $\alpha=1.00$ are not included
    since they are far off and would only clutter the figure.) The data in panel (d) are
    particle rotations $[\omega_z/\gdot]_0$ due to the shearing. These figures suggest a
    crossover to different behaviors at $\alpha\approx1.2$.}
  \label{fig:vs-alpha}
\end{figure}

\paragraph{Summary and discussion}

To summarize, we have shown that a change from spherical to slightly ellipsoidal particles
with aspect ratio $\alpha=1.02$, gives an altogether different microscopic dynamics.
Comparing spheres and ellipsoids close to jamming it is found that the translational
velocity is reduced by 80\%, and that the rotations get a different role for the
ellipsoids, contributing to the particles ability to fit together.  The dramatic
difference is also seen when separating the dissipation into contributions from the
translational and the rotational velocities. For spheres the dissipation is dominated by
the translational motion whereas the dissipation in the ellipsoids is instead dominated by
the rotations. Several quantities show different behaviors above and below $\alpha\approx1.2$,
suggesting the existence of two different regions for ellipsoids with low and high aspect ratio.

What can now be concluded about the relation between this different dynamics and the
quartic modes?  To settle that question beyond possible doubt would require a study where
both the quartic modes and the microscopic dynamics were determined simulataneously, which
is well beyond the scope of the present work. When trying to find an explanation for the
present findings from properties of isotopically jammed systems, it does however seem that
the presence of quartic modes is a natural candidate simply because they are the
outstanding difference between packings of ellipsoids and packings of spheres. What gives
additional credibility to such a conclusion is the observation that this different
dynamics very strongly affects the rotations, which is in agreement with the quartic modes
being primarily rotational in character \cite{Mailman:ellipse-jamming,
  Schreck:ellipsoidal}. We therefore conclude that there is strong evidence for
considering the very different dynamics in shear-driven ellipsoids to be effects of the
quartic vibrational modes on the shear-driven systems.

\begin{acknowledgments}
  We thank S. Teitel for many discussions.  The simulations were performed on resources
  provided by the Swedish National Infrastructure for Computing (SNIC) at HPC2N.
\end{acknowledgments}

\bibliography{j}

\begin{thebibliography}{29}%
\makeatletter
\providecommand \@ifxundefined [1]{%
 \@ifx{#1\undefined}
}%
\providecommand \@ifnum [1]{%
 \ifnum #1\expandafter \@firstoftwo
 \else \expandafter \@secondoftwo
 \fi
}%
\providecommand \@ifx [1]{%
 \ifx #1\expandafter \@firstoftwo
 \else \expandafter \@secondoftwo
 \fi
}%
\providecommand \natexlab [1]{#1}%
\providecommand \enquote  [1]{``#1''}%
\providecommand \bibnamefont  [1]{#1}%
\providecommand \bibfnamefont [1]{#1}%
\providecommand \citenamefont [1]{#1}%
\providecommand \href@noop [0]{\@secondoftwo}%
\providecommand \href [0]{\begingroup \@sanitize@url \@href}%
\providecommand \@href[1]{\@@startlink{#1}\@@href}%
\providecommand \@@href[1]{\endgroup#1\@@endlink}%
\providecommand \@sanitize@url [0]{\catcode `\\12\catcode `\$12\catcode
  `\&12\catcode `\#12\catcode `\^12\catcode `\_12\catcode `\%12\relax}%
\providecommand \@@startlink[1]{}%
\providecommand \@@endlink[0]{}%
\providecommand \url  [0]{\begingroup\@sanitize@url \@url }%
\providecommand \@url [1]{\endgroup\@href {#1}{\urlprefix }}%
\providecommand \urlprefix  [0]{URL }%
\providecommand \Eprint [0]{\href }%
\providecommand \doibase [0]{http://dx.doi.org/}%
\providecommand \selectlanguage [0]{\@gobble}%
\providecommand \bibinfo  [0]{\@secondoftwo}%
\providecommand \bibfield  [0]{\@secondoftwo}%
\providecommand \translation [1]{[#1]}%
\providecommand \BibitemOpen [0]{}%
\providecommand \bibitemStop [0]{}%
\providecommand \bibitemNoStop [0]{.\EOS\space}%
\providecommand \EOS [0]{\spacefactor3000\relax}%
\providecommand \BibitemShut  [1]{\csname bibitem#1\endcsname}%
\let\auto@bib@innerbib\@empty
%</preamble>
\bibitem [{\citenamefont {Donev}\ \emph {et~al.}(2005)\citenamefont {Donev},
  \citenamefont {Torquato},\ and\ \citenamefont
  {Stillinger}}]{Donev_TS:packing}%
  \BibitemOpen
  \bibfield  {author} {\bibinfo {author} {\bibfnamefont {A.}~\bibnamefont
  {Donev}}, \bibinfo {author} {\bibfnamefont {S.}~\bibnamefont {Torquato}}, \
  and\ \bibinfo {author} {\bibfnamefont {F.~H.}\ \bibnamefont {Stillinger}},\
  }\href {\doibase 10.1103/PhysRevE.71.011105} {\bibfield  {journal} {\bibinfo
  {journal} {Phys. Rev. E}\ }\textbf {\bibinfo {volume} {71}},\ \bibinfo
  {pages} {011105} (\bibinfo {year} {2005})}\BibitemShut {NoStop}%
\bibitem [{\citenamefont {Berthier}\ and\ \citenamefont
  {Witten}(2009)}]{Berthier_Witten:PRE2009}%
  \BibitemOpen
  \bibfield  {author} {\bibinfo {author} {\bibfnamefont {L.}~\bibnamefont
  {Berthier}}\ and\ \bibinfo {author} {\bibfnamefont {T.~A.}\ \bibnamefont
  {Witten}},\ }\href {\doibase 10.1103/PhysRevE.80.021502} {\bibfield
  {journal} {\bibinfo  {journal} {Phys. Rev. E}\ }\textbf {\bibinfo {volume}
  {80}},\ \bibinfo {pages} {021502} (\bibinfo {year} {2009})}\BibitemShut
  {NoStop}%
\bibitem [{\citenamefont {V{\aa}gberg}\ \emph
  {et~al.}(2011{\natexlab{a}})\citenamefont {V{\aa}gberg}, \citenamefont
  {Olsson},\ and\ \citenamefont {Teitel}}]{Vagberg_OT:protocol}%
  \BibitemOpen
  \bibfield  {author} {\bibinfo {author} {\bibfnamefont {D.}~\bibnamefont
  {V{\aa}gberg}}, \bibinfo {author} {\bibfnamefont {P.}~\bibnamefont {Olsson}},
  \ and\ \bibinfo {author} {\bibfnamefont {S.}~\bibnamefont {Teitel}},\ }\href
  {\doibase 10.1103/PhysRevE.83.031307} {\bibfield  {journal} {\bibinfo
  {journal} {Phys.\ Rev.\ E}\ }\textbf {\bibinfo {volume} {83}},\ \bibinfo
  {pages} {031307} (\bibinfo {year} {2011}{\natexlab{a}})}\BibitemShut
  {NoStop}%
\bibitem [{\citenamefont {Ozawa}\ \emph {et~al.}(2017)\citenamefont {Ozawa},
  \citenamefont {Berthier},\ and\ \citenamefont {Coslovich}}]{Ozawa:2017}%
  \BibitemOpen
  \bibfield  {author} {\bibinfo {author} {\bibfnamefont {M.}~\bibnamefont
  {Ozawa}}, \bibinfo {author} {\bibfnamefont {L.}~\bibnamefont {Berthier}}, \
  and\ \bibinfo {author} {\bibfnamefont {D.}~\bibnamefont {Coslovich}},\
  }\href@noop {} {\bibfield  {journal} {\bibinfo  {journal} {SciPost Phys.}\
  }\textbf {\bibinfo {volume} {3}},\ \bibinfo {pages} {027} (\bibinfo {year}
  {2017})}\BibitemShut {NoStop}%
\bibitem [{\citenamefont {O'Hern}\ \emph {et~al.}(2002)\citenamefont {O'Hern},
  \citenamefont {Langer}, \citenamefont {Liu},\ and\ \citenamefont
  {Nagel}}]{OHern_Langer_Liu_Nagel:2002}%
  \BibitemOpen
  \bibfield  {author} {\bibinfo {author} {\bibfnamefont {C.~S.}\ \bibnamefont
  {O'Hern}}, \bibinfo {author} {\bibfnamefont {S.~A.}\ \bibnamefont {Langer}},
  \bibinfo {author} {\bibfnamefont {A.~J.}\ \bibnamefont {Liu}}, \ and\
  \bibinfo {author} {\bibfnamefont {S.~R.}\ \bibnamefont {Nagel}},\ }\href@noop
  {} {\bibfield  {journal} {\bibinfo  {journal} {Phys. Rev. Lett.}\ }\textbf
  {\bibinfo {volume} {88}},\ \bibinfo {pages} {075507} (\bibinfo {year}
  {2002})}\BibitemShut {NoStop}%
\bibitem [{\citenamefont {O'Hern}\ \emph {et~al.}(2003)\citenamefont {O'Hern},
  \citenamefont {Silbert}, \citenamefont {Liu},\ and\ \citenamefont
  {Nagel}}]{OHern_Silbert_Liu_Nagel:2003}%
  \BibitemOpen
  \bibfield  {author} {\bibinfo {author} {\bibfnamefont {C.~S.}\ \bibnamefont
  {O'Hern}}, \bibinfo {author} {\bibfnamefont {L.~E.}\ \bibnamefont {Silbert}},
  \bibinfo {author} {\bibfnamefont {A.~J.}\ \bibnamefont {Liu}}, \ and\
  \bibinfo {author} {\bibfnamefont {S.~R.}\ \bibnamefont {Nagel}},\ }\href
  {\doibase 10.1103/PhysRevE.68.011306} {\bibfield  {journal} {\bibinfo
  {journal} {Phys. Rev. E}\ }\textbf {\bibinfo {volume} {68}},\ \bibinfo
  {pages} {011306} (\bibinfo {year} {2003})}\BibitemShut {NoStop}%
\bibitem [{\citenamefont {V{\aa}gberg}\ \emph
  {et~al.}(2011{\natexlab{b}})\citenamefont {V{\aa}gberg}, \citenamefont
  {Valdez-Balderas}, \citenamefont {Moore}, \citenamefont {Olsson},\ and\
  \citenamefont {Teitel}}]{Vagberg_VMOT:jam-fss}%
  \BibitemOpen
  \bibfield  {author} {\bibinfo {author} {\bibfnamefont {D.}~\bibnamefont
  {V{\aa}gberg}}, \bibinfo {author} {\bibfnamefont {D.}~\bibnamefont
  {Valdez-Balderas}}, \bibinfo {author} {\bibfnamefont {M.~A.}\ \bibnamefont
  {Moore}}, \bibinfo {author} {\bibfnamefont {P.}~\bibnamefont {Olsson}}, \
  and\ \bibinfo {author} {\bibfnamefont {S.}~\bibnamefont {Teitel}},\ }\href
  {\doibase 10.1103/PhysRevE.83.030303} {\bibfield  {journal} {\bibinfo
  {journal} {Phys.\ Rev.\ E}\ }\textbf {\bibinfo {volume} {83}},\ \bibinfo
  {pages} {030303(R)} (\bibinfo {year} {2011}{\natexlab{b}})}\BibitemShut
  {NoStop}%
\bibitem [{\citenamefont {Heussinger}\ and\ \citenamefont
  {Barrat}(2009)}]{Heussinger_Barrat:2009}%
  \BibitemOpen
  \bibfield  {author} {\bibinfo {author} {\bibfnamefont {C.}~\bibnamefont
  {Heussinger}}\ and\ \bibinfo {author} {\bibfnamefont {J.-L.}\ \bibnamefont
  {Barrat}},\ }\href {\doibase 10.1103/PhysRevLett.102.218303} {\bibfield
  {journal} {\bibinfo  {journal} {Phys. Rev. Lett.}\ }\textbf {\bibinfo
  {volume} {102}},\ \bibinfo {pages} {218303} (\bibinfo {year}
  {2009})}\BibitemShut {NoStop}%
\bibitem [{\citenamefont {Heussinger}\ \emph {et~al.}(2010)\citenamefont
  {Heussinger}, \citenamefont {Chaudhuri},\ and\ \citenamefont
  {Barrat}}]{Heussinger_Chaudhuri_Barrat-Softmatter}%
  \BibitemOpen
  \bibfield  {author} {\bibinfo {author} {\bibfnamefont {C.}~\bibnamefont
  {Heussinger}}, \bibinfo {author} {\bibfnamefont {P.}~\bibnamefont
  {Chaudhuri}}, \ and\ \bibinfo {author} {\bibfnamefont {J.-L.}\ \bibnamefont
  {Barrat}},\ }\href {\doibase 10.1039/b927228c} {\bibfield  {journal}
  {\bibinfo  {journal} {Soft Matter}\ }\textbf {\bibinfo {volume} {6}},\
  \bibinfo {pages} {3050} (\bibinfo {year} {2010})}\BibitemShut {NoStop}%
\bibitem [{\citenamefont {Olsson}\ and\ \citenamefont
  {Teitel}(2007)}]{Olsson_Teitel:jamming}%
  \BibitemOpen
  \bibfield  {author} {\bibinfo {author} {\bibfnamefont {P.}~\bibnamefont
  {Olsson}}\ and\ \bibinfo {author} {\bibfnamefont {S.}~\bibnamefont
  {Teitel}},\ }\href {\doibase 10.1103/PhysRevLett.99.178001} {\bibfield
  {journal} {\bibinfo  {journal} {Phys. Rev. Lett.}\ }\textbf {\bibinfo
  {volume} {99}},\ \bibinfo {pages} {178001} (\bibinfo {year}
  {2007})}\BibitemShut {NoStop}%
\bibitem [{\citenamefont {Hatano}(2008)}]{Hatano:2008}%
  \BibitemOpen
  \bibfield  {author} {\bibinfo {author} {\bibfnamefont {T.}~\bibnamefont
  {Hatano}},\ }\href@noop {} {\bibfield  {journal} {\bibinfo  {journal} {J.
  Phys. Soc. Jpn.}\ }\textbf {\bibinfo {volume} {77}},\ \bibinfo {pages}
  {123002} (\bibinfo {year} {2008})}\BibitemShut {NoStop}%
\bibitem [{\citenamefont {Hatano}(2011)}]{Hatano:2011}%
  \BibitemOpen
  \bibfield  {author} {\bibinfo {author} {\bibfnamefont {T.}~\bibnamefont
  {Hatano}},\ }\href {\doibase 10.1088/1742-6596/319/1/012011} {\bibfield
  {journal} {\bibinfo  {journal} {Journal of Physics: Conference Series}\
  }\textbf {\bibinfo {volume} {319}},\ \bibinfo {pages} {012011} (\bibinfo
  {year} {2011})}\BibitemShut {NoStop}%
\bibitem [{\citenamefont {Otsuki}\ and\ \citenamefont
  {Hayakawa}(2009)}]{Otsuki_Hayakawa:2009b}%
  \BibitemOpen
  \bibfield  {author} {\bibinfo {author} {\bibfnamefont {M.}~\bibnamefont
  {Otsuki}}\ and\ \bibinfo {author} {\bibfnamefont {H.}~\bibnamefont
  {Hayakawa}},\ }\href {\doibase 10.1103/PhysRevE.80.011308} {\bibfield
  {journal} {\bibinfo  {journal} {Phys. Rev. E}\ }\textbf {\bibinfo {volume}
  {80}},\ \bibinfo {pages} {011308} (\bibinfo {year} {2009})}\BibitemShut
  {NoStop}%
\bibitem [{\citenamefont {Hatano}(2009)}]{Hatano:2009}%
  \BibitemOpen
  \bibfield  {author} {\bibinfo {author} {\bibfnamefont {T.}~\bibnamefont
  {Hatano}},\ }\href {\doibase 10.1103/PhysRevE.79.050301} {\bibfield
  {journal} {\bibinfo  {journal} {Phys. Rev. E}\ }\textbf {\bibinfo {volume}
  {79}},\ \bibinfo {pages} {050301(R)} (\bibinfo {year} {2009})}\BibitemShut
  {NoStop}%
\bibitem [{\citenamefont {Hatano}(2010)}]{Hatano:2010}%
  \BibitemOpen
  \bibfield  {author} {\bibinfo {author} {\bibfnamefont {T.}~\bibnamefont
  {Hatano}},\ }\href {\doibase 10.1143/PTPS.184.143} {\bibfield  {journal}
  {\bibinfo  {journal} {Prog.\ Theor.\ Phys.\ Suppl.}\ }\textbf {\bibinfo
  {volume} {184}},\ \bibinfo {pages} {143} (\bibinfo {year}
  {2010})}\BibitemShut {NoStop}%
\bibitem [{\citenamefont {Tighe}\ \emph {et~al.}(2010)\citenamefont {Tighe},
  \citenamefont {Woldhuis}, \citenamefont {Remmers}, \citenamefont {van
  Saarloos},\ and\ \citenamefont {van Hecke}}]{Tighe_WRvSvH}%
  \BibitemOpen
  \bibfield  {author} {\bibinfo {author} {\bibfnamefont {B.~P.}\ \bibnamefont
  {Tighe}}, \bibinfo {author} {\bibfnamefont {E.}~\bibnamefont {Woldhuis}},
  \bibinfo {author} {\bibfnamefont {J.~J.~C.}\ \bibnamefont {Remmers}},
  \bibinfo {author} {\bibfnamefont {W.}~\bibnamefont {van Saarloos}}, \ and\
  \bibinfo {author} {\bibfnamefont {M.}~\bibnamefont {van Hecke}},\ }\href
  {\doibase 10.1103/PhysRevLett.105.088303} {\bibfield  {journal} {\bibinfo
  {journal} {Phys. Rev. Lett.}\ }\textbf {\bibinfo {volume} {105}},\ \bibinfo
  {pages} {088303} (\bibinfo {year} {2010})}\BibitemShut {NoStop}%
\bibitem [{\citenamefont {Marschall}\ \emph {et~al.}(2019)\citenamefont
  {Marschall}, \citenamefont {Keta}, \citenamefont {Olsson},\ and\
  \citenamefont {Teitel}}]{Marschall_Keta_Olsson_Teitel:2019}%
  \BibitemOpen
  \bibfield  {author} {\bibinfo {author} {\bibfnamefont {T.}~\bibnamefont
  {Marschall}}, \bibinfo {author} {\bibfnamefont {Y.-E.}\ \bibnamefont {Keta}},
  \bibinfo {author} {\bibfnamefont {P.}~\bibnamefont {Olsson}}, \ and\ \bibinfo
  {author} {\bibfnamefont {S.}~\bibnamefont {Teitel}},\ }\href {\doibase
  10.1103/PhysRevLett.122.188002} {\bibfield  {journal} {\bibinfo  {journal}
  {Phys. Rev. Lett.}\ }\textbf {\bibinfo {volume} {122}},\ \bibinfo {pages}
  {188002} (\bibinfo {year} {2019})}\BibitemShut {NoStop}%
\bibitem [{\citenamefont {Marschall}\ and\ \citenamefont
  {Teitel}(2018)}]{Marschall_Teitel:sph-cyl-2018}%
  \BibitemOpen
  \bibfield  {author} {\bibinfo {author} {\bibfnamefont {T.}~\bibnamefont
  {Marschall}}\ and\ \bibinfo {author} {\bibfnamefont {S.}~\bibnamefont
  {Teitel}},\ }\href {\doibase 10.1103/PhysRevE.97.012905} {\bibfield
  {journal} {\bibinfo  {journal} {Phys. Rev. E}\ }\textbf {\bibinfo {volume}
  {97}},\ \bibinfo {pages} {012905} (\bibinfo {year} {2018})}\BibitemShut
  {NoStop}%
\bibitem [{\citenamefont {Marschall}\ and\ \citenamefont
  {Teitel}(2019)}]{Marschall_Teitel:sph-cyl-2019}%
  \BibitemOpen
  \bibfield  {author} {\bibinfo {author} {\bibfnamefont {T.~A.}\ \bibnamefont
  {Marschall}}\ and\ \bibinfo {author} {\bibfnamefont {S.}~\bibnamefont
  {Teitel}},\ }\href {\doibase 10.1103/PhysRevE.100.032906} {\bibfield
  {journal} {\bibinfo  {journal} {Phys. Rev. E}\ }\textbf {\bibinfo {volume}
  {100}},\ \bibinfo {pages} {032906} (\bibinfo {year} {2019})}\BibitemShut
  {NoStop}%
\bibitem [{\citenamefont {Marschall}\ \emph {et~al.}(2020)\citenamefont
  {Marschall}, \citenamefont {Van~Hoesen},\ and\ \citenamefont
  {Teitel}}]{Marschall_Teitel:sph-cyl-2020}%
  \BibitemOpen
  \bibfield  {author} {\bibinfo {author} {\bibfnamefont {T.~A.}\ \bibnamefont
  {Marschall}}, \bibinfo {author} {\bibfnamefont {D.}~\bibnamefont
  {Van~Hoesen}}, \ and\ \bibinfo {author} {\bibfnamefont {S.}~\bibnamefont
  {Teitel}},\ }\href {\doibase 10.1103/PhysRevE.101.032901} {\bibfield
  {journal} {\bibinfo  {journal} {Phys. Rev. E}\ }\textbf {\bibinfo {volume}
  {101}},\ \bibinfo {pages} {032901} (\bibinfo {year} {2020})}\BibitemShut
  {NoStop}%
\bibitem [{\citenamefont {Donev}\ \emph {et~al.}(2004)\citenamefont {Donev},
  \citenamefont {Cisse}, \citenamefont {Sachs}, \citenamefont {Variano},
  \citenamefont {Stillinger}, \citenamefont {Connelly}, \citenamefont
  {Torquato},\ and\ \citenamefont {Chaikin}}]{Donev-Elli:2004}%
  \BibitemOpen
  \bibfield  {author} {\bibinfo {author} {\bibfnamefont {A.}~\bibnamefont
  {Donev}}, \bibinfo {author} {\bibfnamefont {I.}~\bibnamefont {Cisse}},
  \bibinfo {author} {\bibfnamefont {D.}~\bibnamefont {Sachs}}, \bibinfo
  {author} {\bibfnamefont {E.~A.}\ \bibnamefont {Variano}}, \bibinfo {author}
  {\bibfnamefont {F.~H.}\ \bibnamefont {Stillinger}}, \bibinfo {author}
  {\bibfnamefont {R.}~\bibnamefont {Connelly}}, \bibinfo {author}
  {\bibfnamefont {S.}~\bibnamefont {Torquato}}, \ and\ \bibinfo {author}
  {\bibfnamefont {P.~M.}\ \bibnamefont {Chaikin}},\ }\href {\doibase
  10.1126/science.1093010} {\bibfield  {journal} {\bibinfo  {journal}
  {Science}\ }\textbf {\bibinfo {volume} {303}},\ \bibinfo {pages} {990}
  (\bibinfo {year} {2004})}\BibitemShut {NoStop}%
\bibitem [{\citenamefont {Donev}\ \emph {et~al.}(2007)\citenamefont {Donev},
  \citenamefont {Connelly}, \citenamefont {Stillinger},\ and\ \citenamefont
  {Torquato}}]{Donev-Elli:2007}%
  \BibitemOpen
  \bibfield  {author} {\bibinfo {author} {\bibfnamefont {A.}~\bibnamefont
  {Donev}}, \bibinfo {author} {\bibfnamefont {R.}~\bibnamefont {Connelly}},
  \bibinfo {author} {\bibfnamefont {F.~H.}\ \bibnamefont {Stillinger}}, \ and\
  \bibinfo {author} {\bibfnamefont {S.}~\bibnamefont {Torquato}},\ }\href
  {\doibase 0310.1103/PhysRevE.75.051304} {\bibfield  {journal} {\bibinfo
  {journal} {Phys. Rev. E}\ }\textbf {\bibinfo {volume} {75}},\ \bibinfo
  {pages} {051304} (\bibinfo {year} {2007})}\BibitemShut {NoStop}%
\bibitem [{\citenamefont {Mailman}\ \emph {et~al.}(2009)\citenamefont
  {Mailman}, \citenamefont {Schreck}, \citenamefont {O'Hern},\ and\
  \citenamefont {Chakraborty}}]{Mailman:ellipse-jamming}%
  \BibitemOpen
  \bibfield  {author} {\bibinfo {author} {\bibfnamefont {M.}~\bibnamefont
  {Mailman}}, \bibinfo {author} {\bibfnamefont {C.~F.}\ \bibnamefont
  {Schreck}}, \bibinfo {author} {\bibfnamefont {C.~S.}\ \bibnamefont {O'Hern}},
  \ and\ \bibinfo {author} {\bibfnamefont {B.}~\bibnamefont {Chakraborty}},\
  }\href {\doibase 10.1103/PhysRevLett.102.255501} {\bibfield  {journal}
  {\bibinfo  {journal} {Phys. Rev. Lett.}\ }\textbf {\bibinfo {volume} {102}},\
  \bibinfo {pages} {255501} (\bibinfo {year} {2009})}\BibitemShut {NoStop}%
\bibitem [{\citenamefont {Schreck}\ \emph {et~al.}(2012)\citenamefont
  {Schreck}, \citenamefont {Mailman}, \citenamefont {Chakraborty},\ and\
  \citenamefont {O'Hern}}]{Schreck:ellipsoidal}%
  \BibitemOpen
  \bibfield  {author} {\bibinfo {author} {\bibfnamefont {C.~F.}\ \bibnamefont
  {Schreck}}, \bibinfo {author} {\bibfnamefont {M.}~\bibnamefont {Mailman}},
  \bibinfo {author} {\bibfnamefont {B.}~\bibnamefont {Chakraborty}}, \ and\
  \bibinfo {author} {\bibfnamefont {C.~S.}\ \bibnamefont {O'Hern}},\ }\href
  {\doibase 10.1103/PhysRevE.85.061305} {\bibfield  {journal} {\bibinfo
  {journal} {Phys. Rev. E}\ }\textbf {\bibinfo {volume} {85}},\ \bibinfo
  {pages} {061305} (\bibinfo {year} {2012})}\BibitemShut {NoStop}%
\bibitem [{\citenamefont {B\"orzs\"onyi}\ \emph {et~al.}(2012)\citenamefont
  {B\"orzs\"onyi}, \citenamefont {Szab\'o}, \citenamefont {Wegner},
  \citenamefont {Harth}, \citenamefont {T\"or\"ok}, \citenamefont {Somfai},
  \citenamefont {Bien},\ and\ \citenamefont {Stannarius}}]{B-Stannarius:2012}%
  \BibitemOpen
  \bibfield  {author} {\bibinfo {author} {\bibfnamefont {T.}~\bibnamefont
  {B\"orzs\"onyi}}, \bibinfo {author} {\bibfnamefont {B.}~\bibnamefont
  {Szab\'o}}, \bibinfo {author} {\bibfnamefont {S.}~\bibnamefont {Wegner}},
  \bibinfo {author} {\bibfnamefont {K.}~\bibnamefont {Harth}}, \bibinfo
  {author} {\bibfnamefont {J.}~\bibnamefont {T\"or\"ok}}, \bibinfo {author}
  {\bibfnamefont {E.}~\bibnamefont {Somfai}}, \bibinfo {author} {\bibfnamefont
  {T.}~\bibnamefont {Bien}}, \ and\ \bibinfo {author} {\bibfnamefont
  {R.}~\bibnamefont {Stannarius}},\ }\href {\doibase
  10.1103/PhysRevE.86.051304} {\bibfield  {journal} {\bibinfo  {journal} {Phys.
  Rev. E}\ }\textbf {\bibinfo {volume} {86}},\ \bibinfo {pages} {051304}
  (\bibinfo {year} {2012})}\BibitemShut {NoStop}%
\bibitem [{\citenamefont {Nagy}\ \emph {et~al.}(2017)\citenamefont {Nagy},
  \citenamefont {Claudin}, \citenamefont {B\"orzs\"onyi},\ and\ \citenamefont
  {Somfai}}]{Nagy:2017}%
  \BibitemOpen
  \bibfield  {author} {\bibinfo {author} {\bibfnamefont {D.~B.}\ \bibnamefont
  {Nagy}}, \bibinfo {author} {\bibfnamefont {P.}~\bibnamefont {Claudin}},
  \bibinfo {author} {\bibfnamefont {T.}~\bibnamefont {B\"orzs\"onyi}}, \ and\
  \bibinfo {author} {\bibfnamefont {E.}~\bibnamefont {Somfai}},\ }\href
  {\doibase 10.1103/PhysRevE.96.062903} {\bibfield  {journal} {\bibinfo
  {journal} {Phys. Rev. E}\ }\textbf {\bibinfo {volume} {96}},\ \bibinfo
  {pages} {062903} (\bibinfo {year} {2017})}\BibitemShut {NoStop}%
\bibitem [{\citenamefont {Trulsson}(2018)}]{Trulsson:2018}%
  \BibitemOpen
  \bibfield  {author} {\bibinfo {author} {\bibfnamefont {M.}~\bibnamefont
  {Trulsson}},\ }\href {\doibase 10.1017/jfm.2018.420} {\bibfield  {journal}
  {\bibinfo  {journal} {Journal of Fluid Mechanics}\ }\textbf {\bibinfo
  {volume} {849}},\ \bibinfo {pages} {718–740} (\bibinfo {year}
  {2018})}\BibitemShut {NoStop}%
\bibitem [{\citenamefont {Ikeda}\ \emph {et~al.}(2020)\citenamefont {Ikeda},
  \citenamefont {Brito},\ and\ \citenamefont {Wyart}}]{Ikeda:2020}%
  \BibitemOpen
  \bibfield  {author} {\bibinfo {author} {\bibfnamefont {H.}~\bibnamefont
  {Ikeda}}, \bibinfo {author} {\bibfnamefont {C.}~\bibnamefont {Brito}}, \ and\
  \bibinfo {author} {\bibfnamefont {M.}~\bibnamefont {Wyart}},\ }\href
  {\doibase 10.1088/1742-5468/ab74cb} {\bibfield  {journal} {\bibinfo
  {journal} {Journal of Statistical Mechanics: Theory and Experiment}\ }\textbf
  {\bibinfo {volume} {2020}},\ \bibinfo {pages} {033302} (\bibinfo {year}
  {2020})}\BibitemShut {NoStop}%
\bibitem [{\citenamefont {Olsson}(2019)}]{Olsson:jam-3D}%
  \BibitemOpen
  \bibfield  {author} {\bibinfo {author} {\bibfnamefont {P.}~\bibnamefont
  {Olsson}},\ }\href {\doibase 10.1103/PhysRevLett.122.108003} {\bibfield
  {journal} {\bibinfo  {journal} {Phys. Rev. Lett.}\ }\textbf {\bibinfo
  {volume} {122}},\ \bibinfo {pages} {108003} (\bibinfo {year}
  {2019})}\BibitemShut {NoStop}%
\end{thebibliography}%


\begin{thebibliography}{3}%
\makeatletter
\providecommand \@ifxundefined [1]{%
 \@ifx{#1\undefined}
}%
\providecommand \@ifnum [1]{%
 \ifnum #1\expandafter \@firstoftwo
 \else \expandafter \@secondoftwo
 \fi
}%
\providecommand \@ifx [1]{%
 \ifx #1\expandafter \@firstoftwo
 \else \expandafter \@secondoftwo
 \fi
}%
\providecommand \natexlab [1]{#1}%
\providecommand \enquote  [1]{``#1''}%
\providecommand \bibnamefont  [1]{#1}%
\providecommand \bibfnamefont [1]{#1}%
\providecommand \citenamefont [1]{#1}%
\providecommand \href@noop [0]{\@secondoftwo}%
\providecommand \href [0]{\begingroup \@sanitize@url \@href}%
\providecommand \@href[1]{\@@startlink{#1}\@@href}%
\providecommand \@@href[1]{\endgroup#1\@@endlink}%
\providecommand \@sanitize@url [0]{\catcode `\\12\catcode `\$12\catcode
  `\&12\catcode `\#12\catcode `\^12\catcode `\_12\catcode `\%12\relax}%
\providecommand \@@startlink[1]{}%
\providecommand \@@endlink[0]{}%
\providecommand \url  [0]{\begingroup\@sanitize@url \@url }%
\providecommand \@url [1]{\endgroup\@href {#1}{\urlprefix }}%
\providecommand \urlprefix  [0]{URL }%
\providecommand \Eprint [0]{\href }%
\providecommand \doibase [0]{http://dx.doi.org/}%
\providecommand \selectlanguage [0]{\@gobble}%
\providecommand \bibinfo  [0]{\@secondoftwo}%
\providecommand \bibfield  [0]{\@secondoftwo}%
\providecommand \translation [1]{[#1]}%
\providecommand \BibitemOpen [0]{}%
\providecommand \bibitemStop [0]{}%
\providecommand \bibitemNoStop [0]{.\EOS\space}%
\providecommand \EOS [0]{\spacefactor3000\relax}%
\providecommand \BibitemShut  [1]{\csname bibitem#1\endcsname}%
\let\auto@bib@innerbib\@empty
%</preamble>
\bibitem [{\citenamefont {V\aa{}gberg}\ \emph {et~al.}(2017)\citenamefont
  {V\aa{}gberg}, \citenamefont {Olsson},\ and\ \citenamefont
  {Teitel}}]{Vagberg_Olsson_Teitel:CDrot}%
  \BibitemOpen
  \bibfield  {author} {\bibinfo {author} {\bibfnamefont {D.}~\bibnamefont
  {V\aa{}gberg}}, \bibinfo {author} {\bibfnamefont {P.}~\bibnamefont {Olsson}},
  \ and\ \bibinfo {author} {\bibfnamefont {S.}~\bibnamefont {Teitel}},\ }\href
  {\doibase 10.1103/PhysRevE.95.052903} {\bibfield  {journal} {\bibinfo
  {journal} {Phys. Rev. E}\ }\textbf {\bibinfo {volume} {95}},\ \bibinfo
  {pages} {052903} (\bibinfo {year} {2017})}\BibitemShut {NoStop}%
\bibitem [{\citenamefont {Marschall}\ and\ \citenamefont
  {Teitel}(2019)}]{Marschall_Teitel:sph-cyl-2019}%
  \BibitemOpen
  \bibfield  {author} {\bibinfo {author} {\bibfnamefont {T.~A.}\ \bibnamefont
  {Marschall}}\ and\ \bibinfo {author} {\bibfnamefont {S.}~\bibnamefont
  {Teitel}},\ }\href {\doibase 10.1103/PhysRevE.100.032906} {\bibfield
  {journal} {\bibinfo  {journal} {Phys. Rev. E}\ }\textbf {\bibinfo {volume}
  {100}},\ \bibinfo {pages} {032906} (\bibinfo {year} {2019})}\BibitemShut
  {NoStop}%
\bibitem [{\citenamefont {Ono}\ \emph {et~al.}(2003)\citenamefont {Ono},
  \citenamefont {Tewari}, \citenamefont {Langer},\ and\ \citenamefont
  {Liu}}]{Ono_Tewari_Langer_Liu}%
  \BibitemOpen
  \bibfield  {author} {\bibinfo {author} {\bibfnamefont {I.~K.}\ \bibnamefont
  {Ono}}, \bibinfo {author} {\bibfnamefont {S.}~\bibnamefont {Tewari}},
  \bibinfo {author} {\bibfnamefont {S.~A.}\ \bibnamefont {Langer}}, \ and\
  \bibinfo {author} {\bibfnamefont {A.~J.}\ \bibnamefont {Liu}},\ }\href@noop
  {} {\bibfield  {journal} {\bibinfo  {journal} {Phys. Rev. E}\ }\textbf
  {\bibinfo {volume} {67}},\ \bibinfo {pages} {061503} (\bibinfo {year}
  {2003})}\BibitemShut {NoStop}%
\end{thebibliography}%
\bibliographystyle{apsrev4-1}

\end{document}

\title{Supplemental Material to ``Translational and rotational velocities in shear-driven jamming of ellipsoidal particles''}
\author{Yann-Edwin Keta}
\author{Peter Olsson}
\date{\today}

\maketitle

Note that figures and equations in this Supplemental Material are prefixed with an
``S'', e.g.\ ``Fig.~S1''. ``Fig.~1'', on the other hand, is a figure in the Letter.

\section{General considerations}

\renewcommand{\thefigure}{S\arabic{figure}}
\setcounter{figure}{0}
\renewcommand{\theequation}{S\arabic{equation}}
\setcounter{equation}{0}

\subsection{Ellipses and the quartic modes}

\begin{figure}[h]
  \includegraphics[width=7cm]{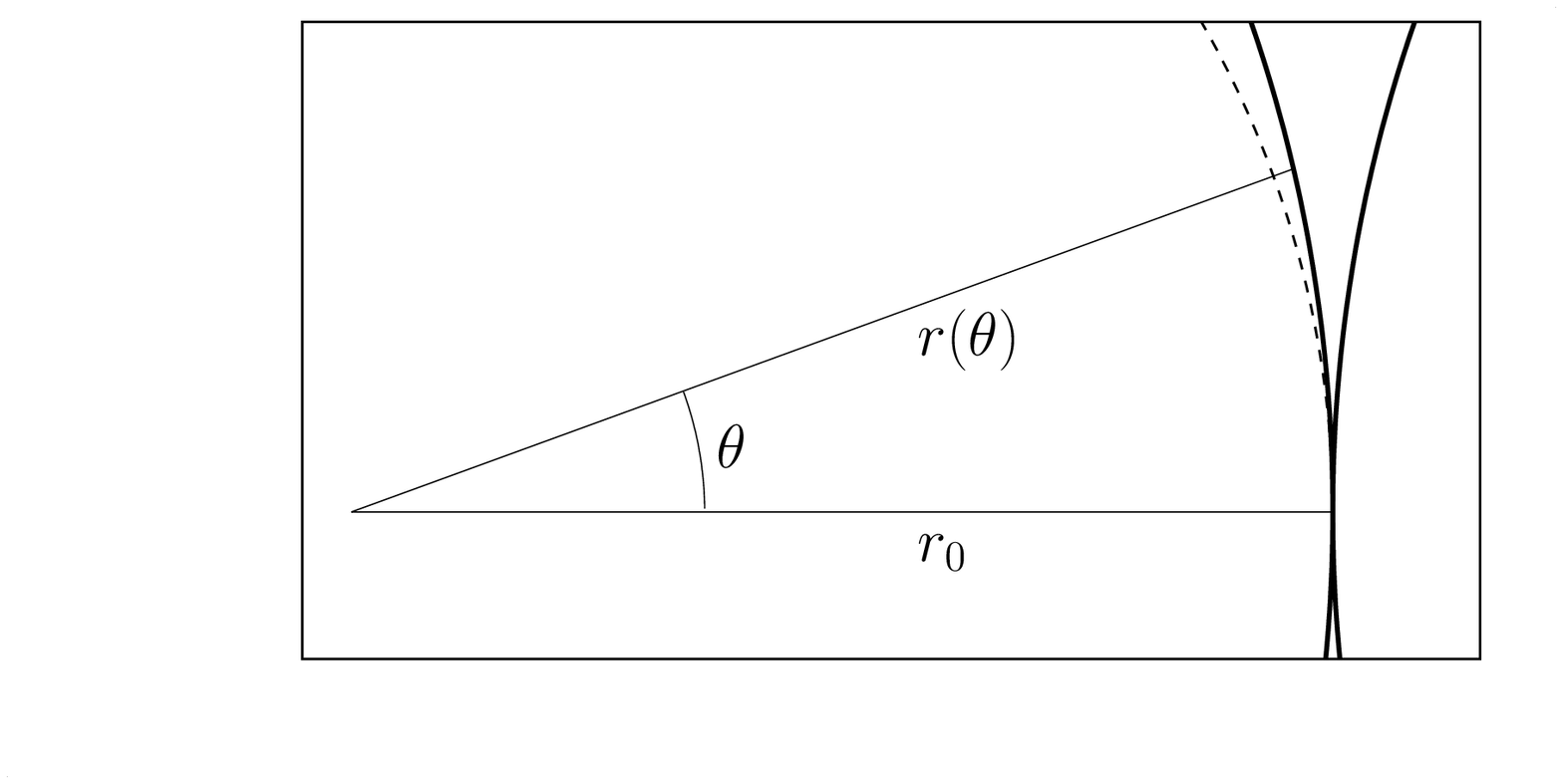}
  \caption{Figure for the discussion of ellipses and the quartic modes.}
  \label{fig:ellipse}
\end{figure}

\Figure{ellipse} shows a closeup of two ellipses, $x^2+a^2y^2=r_0^2$, with $a<1$, touching
at their waists. The dashed curve is from a circle with radius $r_0$ and we here detail
the argument in the main text that the relative distance between the circle and the
ellipsoid depends quadratically on the angle $\theta$, $r(\theta)/r_0-1\sim\theta^2$.

To show this in more detail we note that $x^2=r_0^2-a^2y^2$  in $r^2(\theta) = x^2 + y^2$ becomes
\begin{displaymath}
  r^2(\theta) = r_0^2 + (1-a^2) y^2.
\end{displaymath}
For a small $\theta$ such that $y=r_0\theta$ this becomes
\begin{displaymath}
  r^2(\theta)/r_0^2 = 1 + (1-a^2) \theta^2,
\end{displaymath}
which finally gives
\begin{equation}
  \label{eq:ellipse}
  r(\theta)/r_0 \approx 1 + \frac 1 2 (1-a^2) \theta^2.
\end{equation}

\subsection{Limiting the tangential dissipation}

Our model is a generalization of the model in
Ref.~\onlinecite{Vagberg_Olsson_Teitel:CDrot} to non-circular particles. This model is one
of the simplest possible that includes both contact dissipation and rotation. An undesired
feature of this model is that the tangential part of the dissipation at a given contact
may jump discontinuously to zero when the particles lose contact. With a friction
parameter $\mu$ that controls the maximum size of the dissipation,
$F_t^\mathrm{dis} \leq \mu F^\mathrm{el}$, our model may be described as having
$\mu=\infty$ since any contact, however weak, will be able to sustain an arbitrarily big
dissipative force, given by the velocity difference at the point of contact.

\begin{figure}
  \includegraphics[width=7cm]{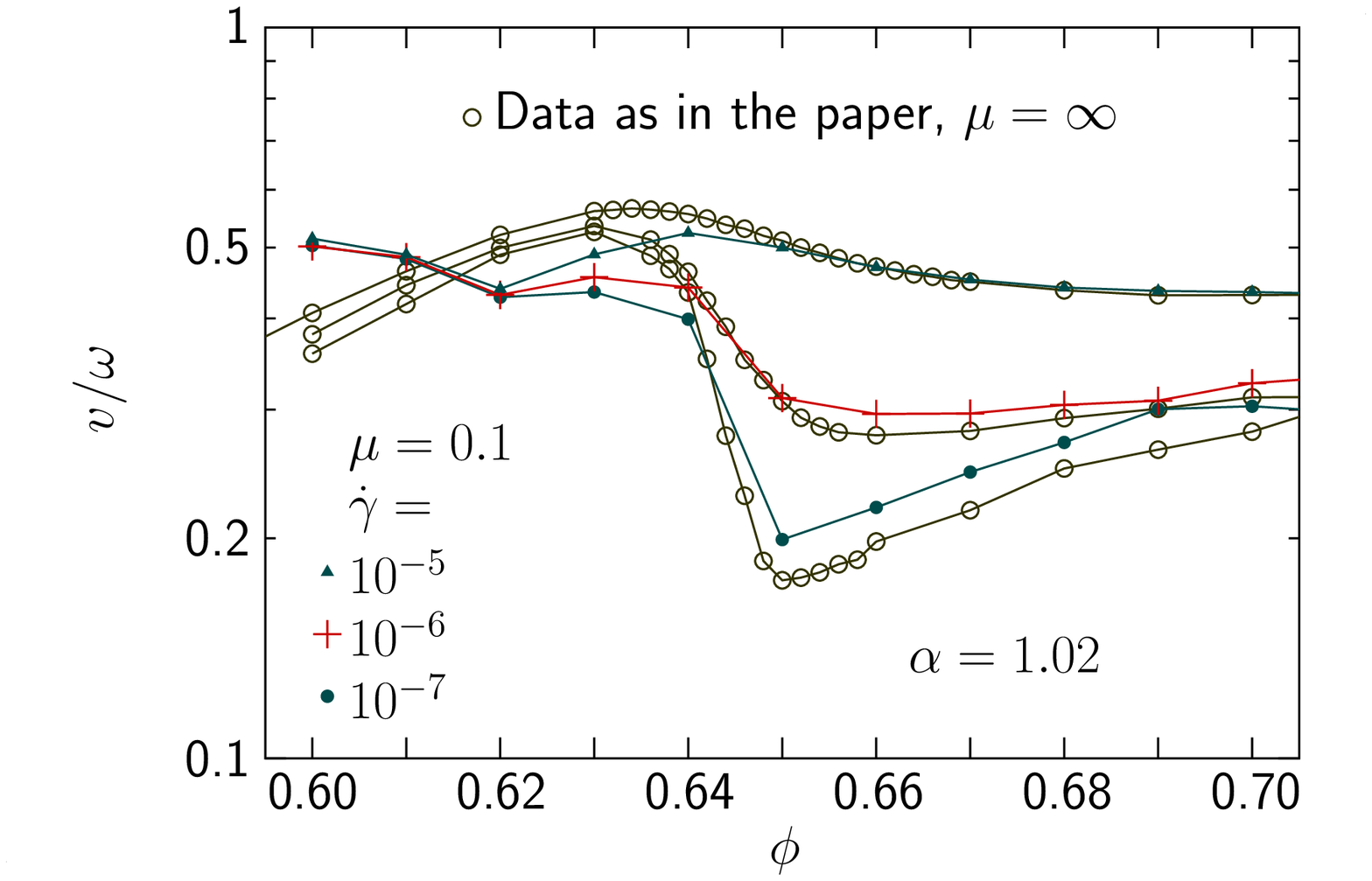}
  \caption{The effect of limiting the tangential dissipation. }
  \label{fig:limtdiss}
\end{figure}

Since the model with $\mu=\infty$ can be argued to be unphysical we have performed
additional simulations with finite $\mu$. It was then found necessary to take a value as
low as $\mu=0.1$ to get any clearly visible changes. \Fig{limtdiss} is therefore a
comparison of $v/\omega$ obtained with $\mu=0.1$ and the data in the paper, which are for
$\mu=\infty$. Even with such a small $\mu$ there are no significant differences at the
density of interest in our studies, which is around $\phi=0.65$. The only notable
difference is a dip in $v/\omega$ at $\phi=0.62$, which is well outside of our region of
interest.  Looking into the reason for this weak dependence on $\mu$ one finds that the
tangential dissipative forces, are typically much smaller than the elastic forces, at the
densities of interest, which means that taking $\mu=0.1$ only affects a small fraction of
contacts with the weakest elastic forces.  We therefore conclude that the results and the
analyses of the present paper would not be significantly affected by instead using a model
that couples the elastic and the dissipative forces.

\subsection{Slightly aspherical particles}
\label{sec:slightly}

\begin{figure}[h]
  % plo compare-6480-100-and-6520-102-all.plo
  \includegraphics[width=7cm]{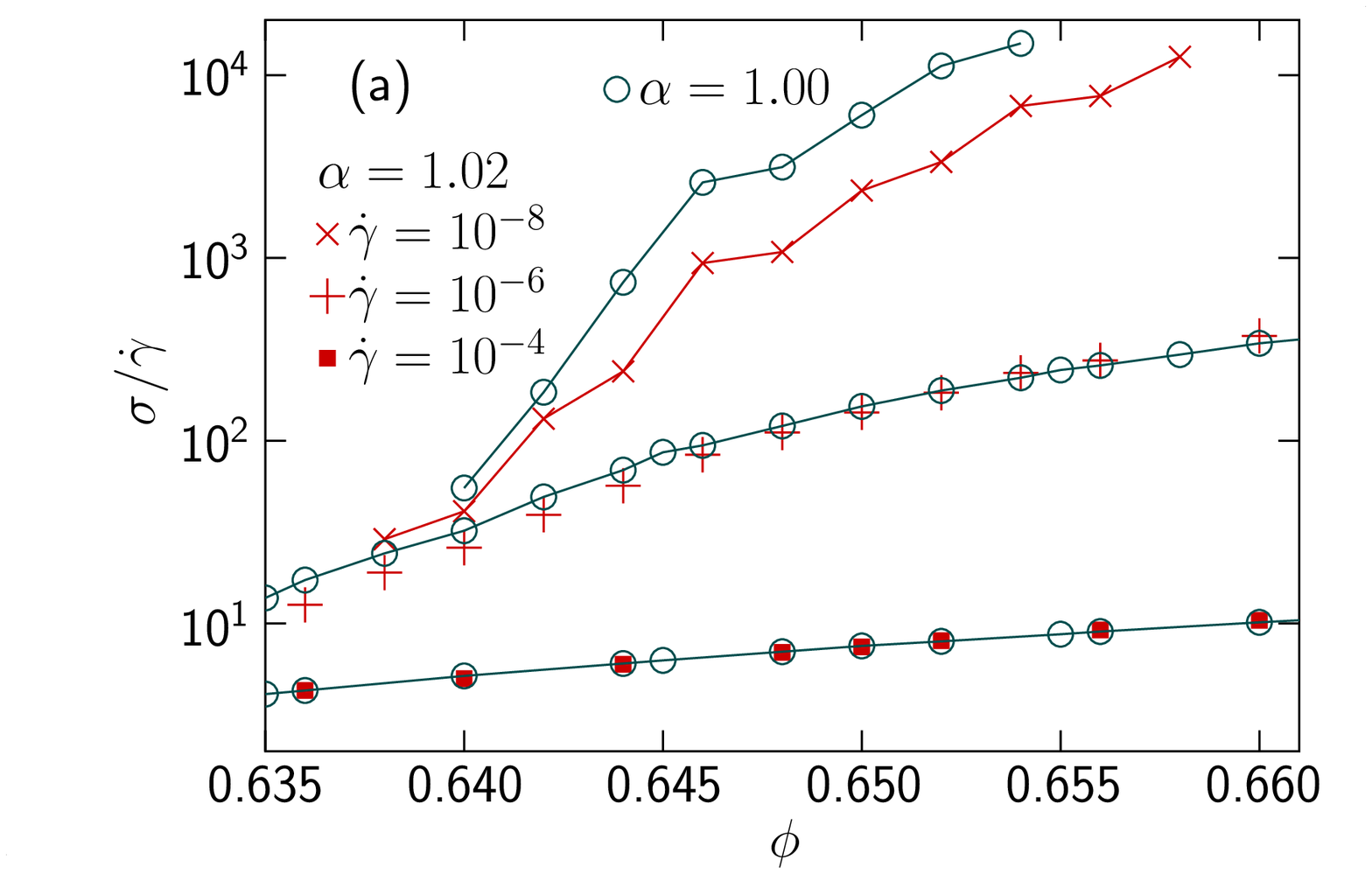}
  \includegraphics[width=7cm]{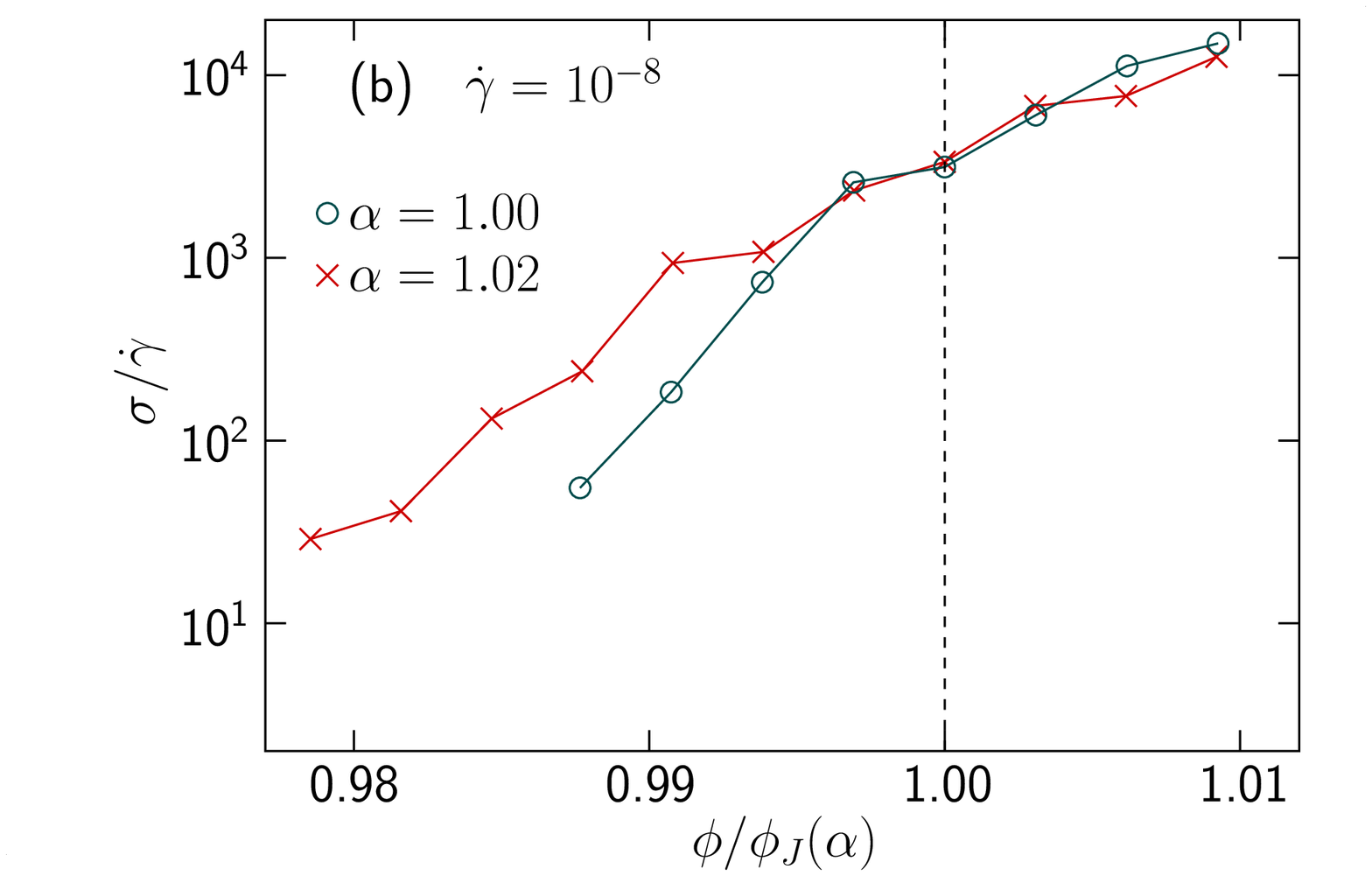}
  \caption{Shear viscosity for spheres, $\alpha=1.00$ and ellipsoids,
    $\alpha=1.02$. Panel~(a) shows direct comparisons for shear strain rates
    $\gdot=10^{-8}$, $10^{-6}$, and $10^{-4}$, which show that a small asphericity only
    has significant effects at low shear strain rates. Panel~(b), which is the same data
    plotted vs $\phi/\phi_J(\alpha)$ where $\phi_J(1.00)=0.648$ and $\phi_J(1.02)=0.652$,
    shows that the change in $\sigma/\gdot$ at small $\gdot$ can be understood as an
    effect of a shift in jamming density. (We expect $\phi_J(\alpha)$ to be close to
    $\phi_J(\alpha)$.) This has also been found to be the case for spherocylinders in 2D
    \cite{Marschall_Teitel:sph-cyl-2019}.}
  \label{fig:S:sigma-compare}
\end{figure}

\Figure{S:sigma-compare} is shear viscosity for spheres and slightly aspherical particles,
$\alpha=1.02$, for different shear strain rates, $\gdot$. The figure shows that these two
cases behave essentially the same for high $\gdot$. This is also reasonable since the
typical particle overlaps at contacts are then of about the same size as typical
deviations of the ellipsoidal particles from the spherical shape, which means that the
effect of the asphericity should be very limited. At lower $\gdot$ the particle overlaps
are smaller and the behavior is also quite different. The shear viscosity of the
ellipsoids at lower $\gdot$ is somewhat lower than for spheres, which is consistent with
the jamming transition taking place at a higher density. \Fig{S:sigma-compare}(b) which
displays this data vs $\phi/\phi_J(\alpha)$, shows that a rescaling of the density can
accomodate this change. Note however that this similarity is not the full story. The main
message of the Letter is that there are big differences in the microscopic dynamics in
spite of the small differences in certain macroscopic quantities.

\section{Translational and rotational velocities}

\begin{figure}[h]
  \includegraphics[bb=50 324 360 657, width=4.2cm]{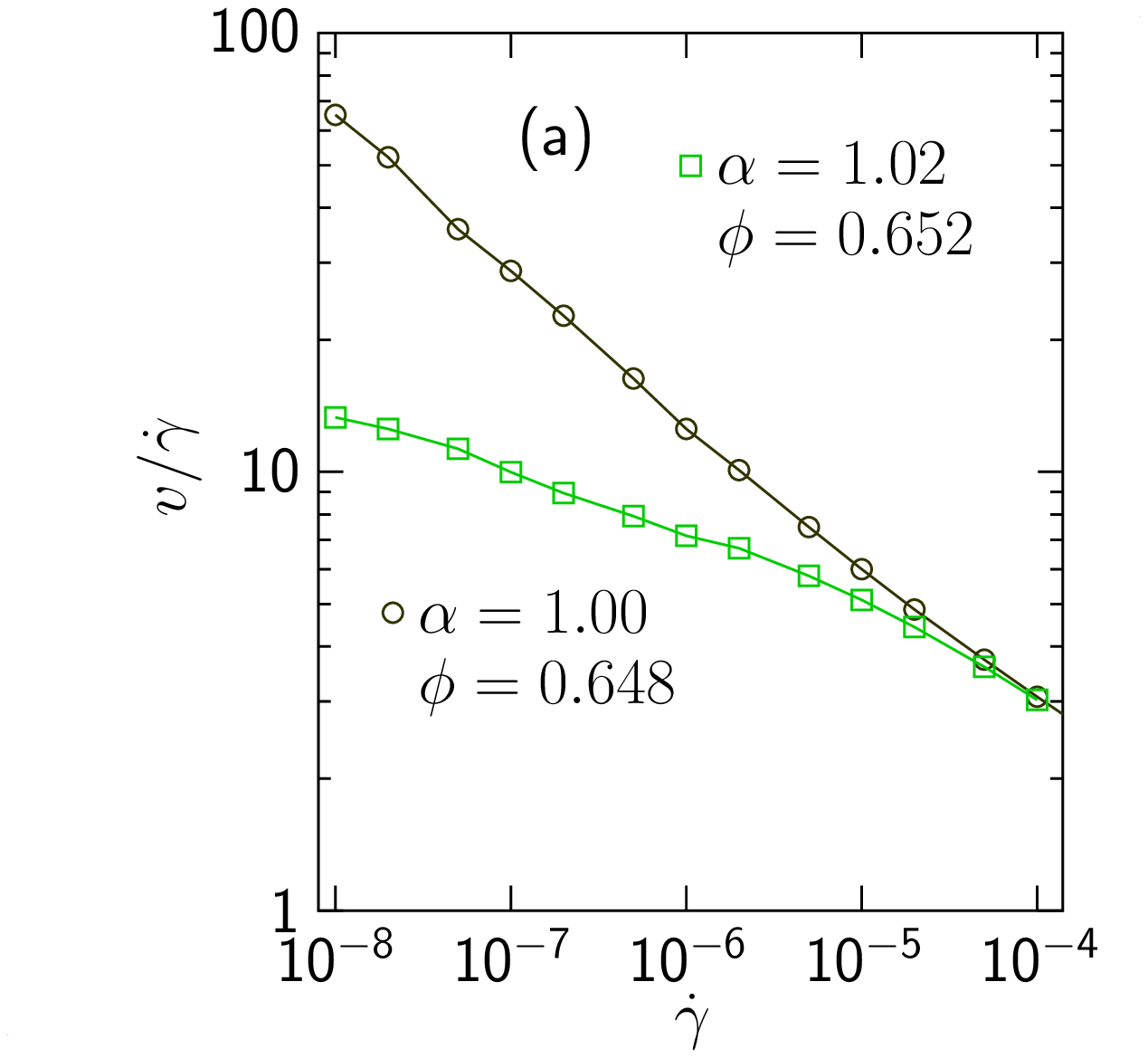}
  \includegraphics[bb=50 324 360 657, width=4.2cm]{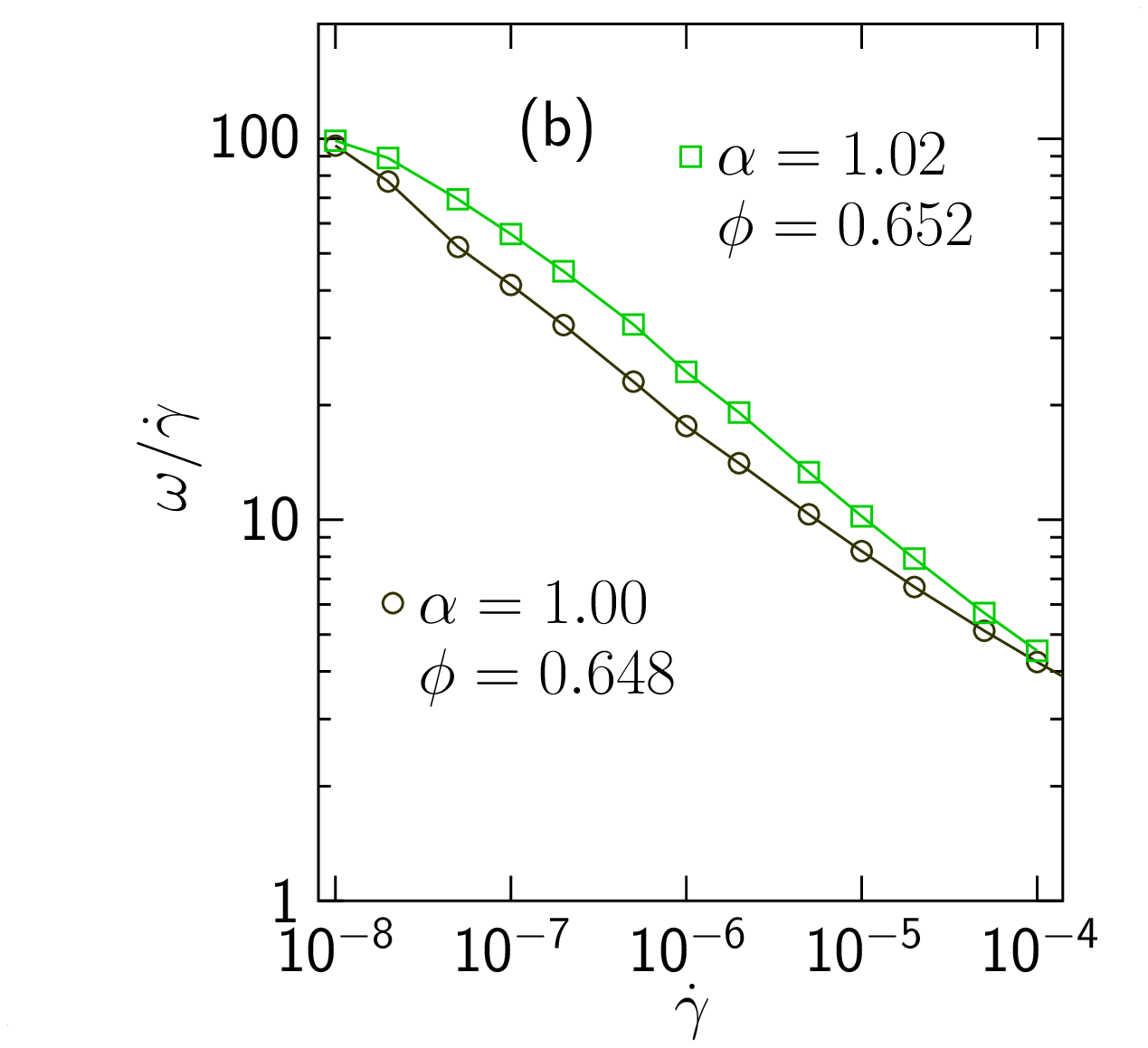}
  \caption{Translational and rotational velocities for ellipsoidal and spherical
    particles. The data are shown at their respective $\phi_J$, $\phi_J(1.00)=0.648$ and
    $\phi_J(1.02)=0.652$. Panel (a) shows that the non-affine translational velocity at
    low shear strain rates is much lower for ellipsoids than for spheres. The rotational
    velocity, panel (b), is in contrast somewhat higher for ellipsoids than for
    spheres. These behaviors of $v$ and $\omega$ together give a dramatic decrease in
    $v/\omega$, shown in \Figs{R}(b) and (c).}
  \label{fig:S:v,omega}
\end{figure}

To look into the origin of the dip in $v/\omega$ in \Fig{R}(b) we now examine $v$ and
$\omega$ for spheres and ellipsoids with $\alpha=1.02$ at the densities
$\phi_J(1.00)=0.648$, and $\phi_J(1.02)=0.652$. \Fig{S:v,omega}(a) and (b) show $v/\gdot$
and $\omega/\gdot$ and we first note that these quantities---open circles in panels (a)
and (b)---behave the same for $\alpha=1.00$, which is consistent with $v/\omega$ being
essentially independent of $\gdot$.  In comparison, for $\alpha=1.02$---open squares in
panels (a) and (b)---$v/\gdot$ increases more slowly than $\omega/\gdot$, which explains
the decrease of $v/\omega$.  The figure shows clearly that it is the lower translational
velocity which is the main reason for the small $v/\omega$ of the ellipsoids.

\begin{figure}
  % plo tot-6480-100-and-6520-102.plo
  \includegraphics[bb=50 324 360 657, width=4.2cm]{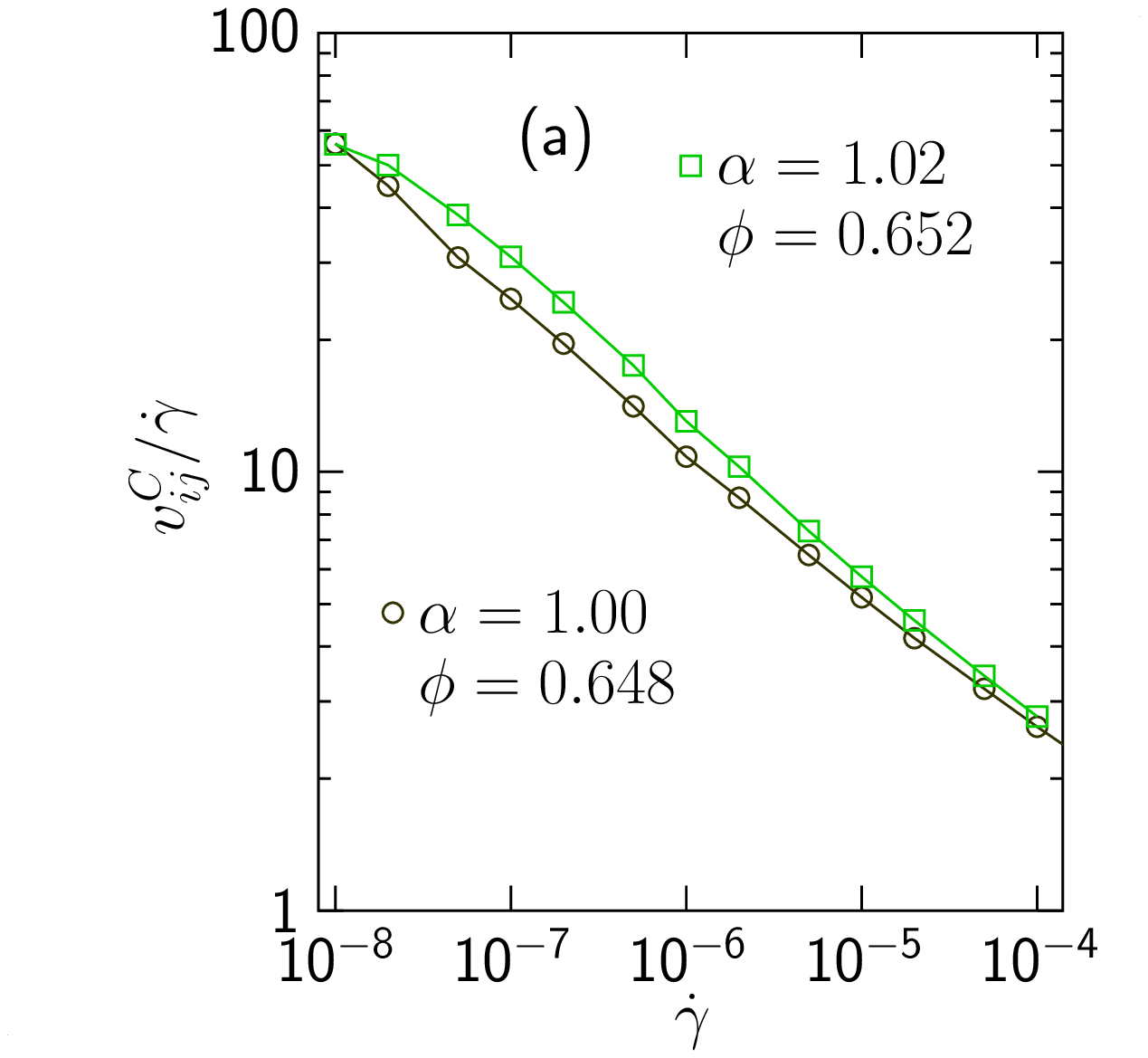}
  % plo R102=6480 plo tot-6480-100-and-6520-102.plo
  \includegraphics[bb=50 324 360 657, width=4.2cm]{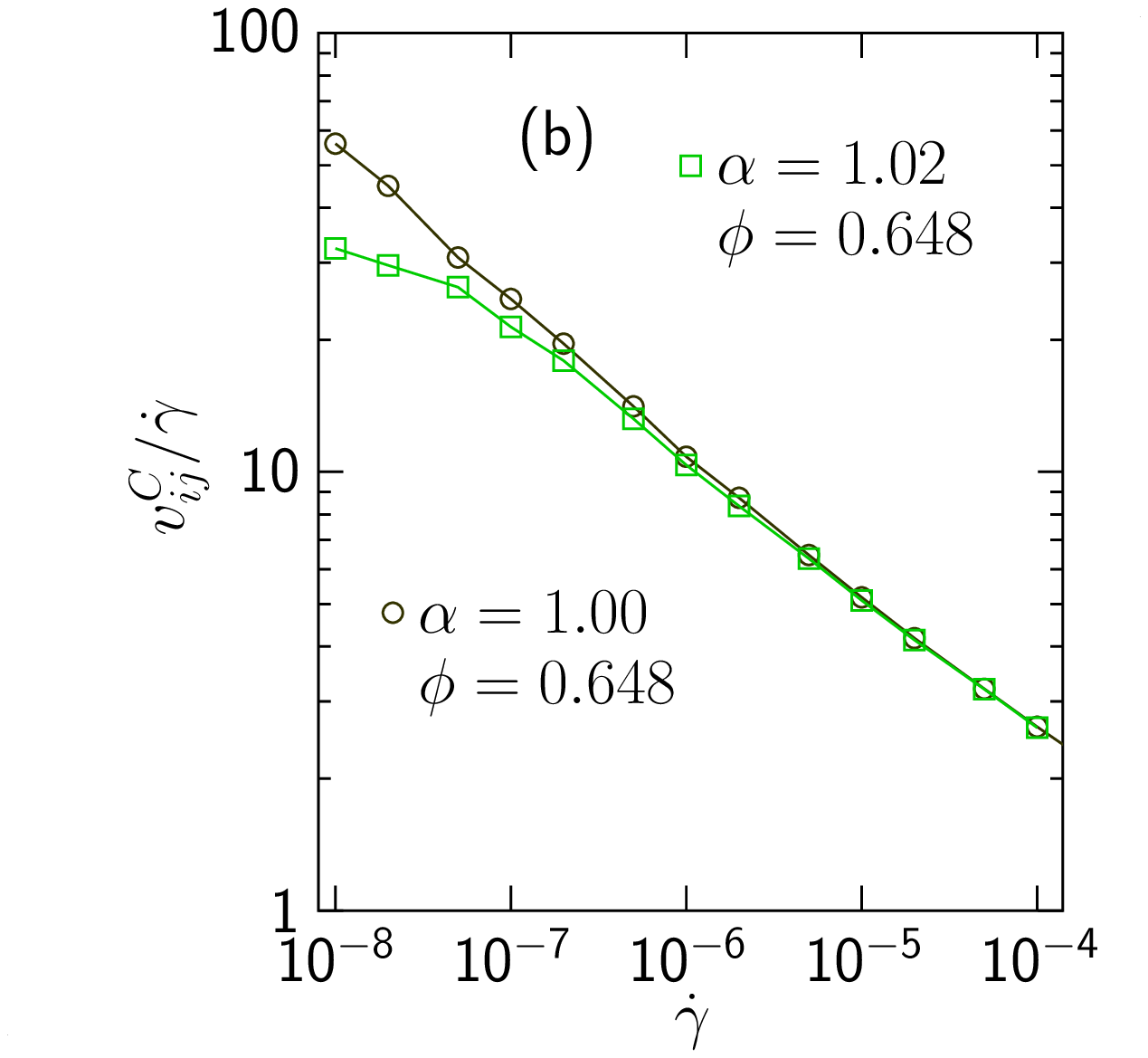}
  % plo R102=6520 plo tot-6480-100-and-6520-102.plo
  \caption{Velocity difference at the point of contact. This is a quantity which is
    related to the dissipation and thereby to the shear viscosity. Panel (a) shows the
    behavior of spheres and ellipsoids at their respective $\phi_J$. Panel (b) which is
    data at the same $\phi$ for both cases is included to show that the ellipsoids, at a
    given density, have lower contact velocities than the spheres, and thereby also a
    lower shear viscosity.}
  \label{fig:S:vCij-gdot}
\end{figure}

The velocities in \Figure{S:v,omega} are single-particle quantities. To complement this,
\Fig{S:vCij-gdot}(a) shows the root-mean-square velocity difference at contact,
$v^C_{ij}$, which can be argued to be a physically more relevant quantity than the
single-particle velocities due to the relation to shear viscosity from power balance,
$Nk_d (z/2) (v^C_{ij})^2=\sigma\gdot V$ \cite{Ono_Tewari_Langer_Liu}. Panel (a) shows
these quantities at their respective $\phi_J(\alpha)$.  $v^C_{ij}$, and thus also the
shear viscosity, is found to be slightly bigger for ellipsoids than for spheres. (The
contact number $z$, which enters the relation to the shear viscosity, is also different
for the two cases, but these differences are too small to significantly affect the
comparison.) \Fig{S:vCij-gdot}(b) is the same quantity at the same density, $\phi=0.648$,
for comparison, which shows that the shear viscosity at the same density and small
$\gdot$, is indeed lower for the ellipsoids, consistent with expectations since
$\phi_J(1.02)>\phi_J(1.00)$.

\section{Translational and rotational contributions to the dissipation}

\begin{figure}
  % plo grc-show.plo
  \includegraphics[width=7cm]{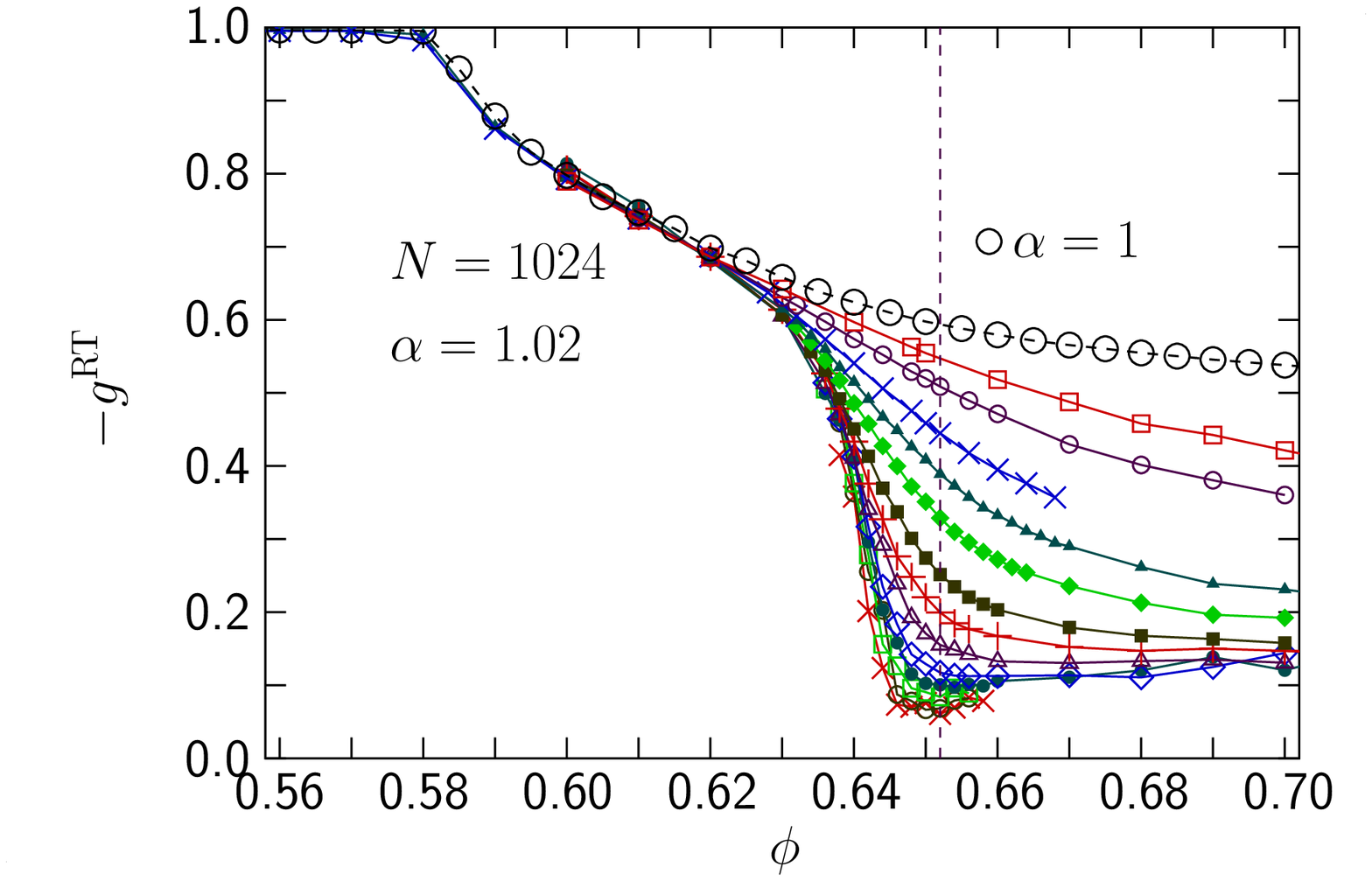}
  \caption{Correlations between different contributions to the contact velocity difference
    for both spheres (one set of open circles) and ellipsoids for several different shear
    rates. The figure shows the (anti-)correlations between the translational contribution
    $\v_{ij}$ and the rotational contribution $\v^R_{ij}$ to the velocity difference at
    contacts. It is found that $\v_{ij}$ and $\v^R_{ij}$ are almost perfectly
    anti-correlated at low densities, $\phi<0.58$, which means that the rotations almost
    entirely compensate for the translational velocity differences. For the spherical
    particles $\v_{ij}$ and $\v^R_{ij}$ remain strongly anti-correlated but for the
    aspherical particles at the lowest shear rates, these correlations almost vanish and
    the two contributions are almost independent.}
  \label{fig:S:gRT}
\end{figure}

\Fig{vijRT} shows an analysis of the dissipation at contacting particles which was done by
splitting up the velocity difference on translational and rotational velocity. We here
make use of the same data to extract information in a different way.

The quantity in focus is the correlation coefficient, which is obtained from the second
term in \Eq{vCij2}, but with a different normalization,
\begin{equation}
  \label{eq:gRT}
  \gRT = \frac{\expt{\v_{ij} \cdot \v^R_{ij}}}{v_{ij}\; v^R_{ij}}.
\end{equation}
Here $v_{ij}$ and $v^R_{ij}$ are the root-mean-square values.

\Fig{S:gRT} shows that both spheres and ellipsoids have a region with almost perfect
anticorrelation, $\gRT\approx -0.996$, for $\phi<0.58$. (Note the different density range;
$\phi$ in \Fig{vijRT} only extends down to $\phi=0.60$.) In this region there is Bagnold
scaling, $p\sim\gdot^2$, and the particles have on average less than two contacts, which
means that rotations can almost altogether compensate for the translational velocity
differences. The behavior then jumps to a region with Newtonian behavior, $p\sim\gdot$ and
a larger number of contacts, $z>4$. It is then no longer possible for the particles to
eliminate the velocity differences by rotations, and the anti-correlation is weaker, down
to $\gRT\approx -0.6$ around the jamming density. Just as in the other quantities
discussed above, there is essentially no $\gdot$-dependence for the spherical particles,
which are shown for $\gdot=10^{-5}$, only.

In \Ref{Vagberg_Olsson_Teitel:CDrot}, Fig.~21, this correlation was explored in a 2D model
by means of a scatter plot which suggested an essentially perfect anticorrelation at low
densities (panel (a)) changing to a weaker anti-correlation at higher densities (panel
(b)). (In the notation of \Ref{Vagberg_Olsson_Teitel:CDrot} $V_{ij,T}$ is the tangential
component of the translational velocity difference, which is almost the same as the full
translational velocity difference, here denoted by $\v_{ij}$.)

As jamming is approached the behavior of ellipsoids again differs considerably from the
behavior of spheres; the correlation coefficient is dropping dramatically. An
extrapolation to the $\gdot\to0$ limit by fitting $\gRT$ at $\phi_J$ to an algebraic
behavior plus a constant, gives $-\gRT_0=0.045$ which implies that $\v_{ij}$ and
$\v^R_{ij}$ are almost entirely decoupled from each other, in sharp contrast to the
essentially perfect anticorrelation at low densities.

\section{Analyses related to Fig.~3}

\subsection{Determinations of $\phi_J(\alpha)$}

As mentioned in the Letter the determination of $\phi_J(\alpha)$ is a truly difficult task
and this is especially so for small asphericities. For examining the dependence on
$\alpha$ it is however necessary to compare behaviors at their respective $\phi_J(\alpha)$
and we have therefore had to resort to some approximate determinations. Since the
quantities in focus do not depend very sensitively on $\phi$, our conclusions do not depend
on the precise values of these $\phi_J(\alpha)$.

In shear driven jamming the jamming density is the density where pressure and shear stress
at small $\gdot$ decay algebraically with $\gdot$. This simple recipe is however difficult
to use due to both corrections to scaling, finite size effects, and---especially for the
case of small asphericities---a crossover from spherical behavior at larger shear rates to
the true ellipsoid behavior, as illustrated in Sec.~\ref{sec:slightly} above. Reliable
determinations of $\phi_J$ need to take all these effects into account.

Since a precise determination of $\phi_J(\alpha)$ is not possible at the present time we
chose to neglect all these complications and instead determine approximate values of
$\phi_J(\alpha)$ as the densities where $p(\gdot; \alpha)$ behave similarly to
$p(\gdot; 1.00)$.  This does not imply the assumption that the models are in the same
universality class, but only that the exponent $q$ in $p\sim\gdot^q$ for ellipsoids is not
altogether different from the value for spheres.

\begin{figure}
  \includegraphics[bb=50 324 360 657, width=4.2cm]{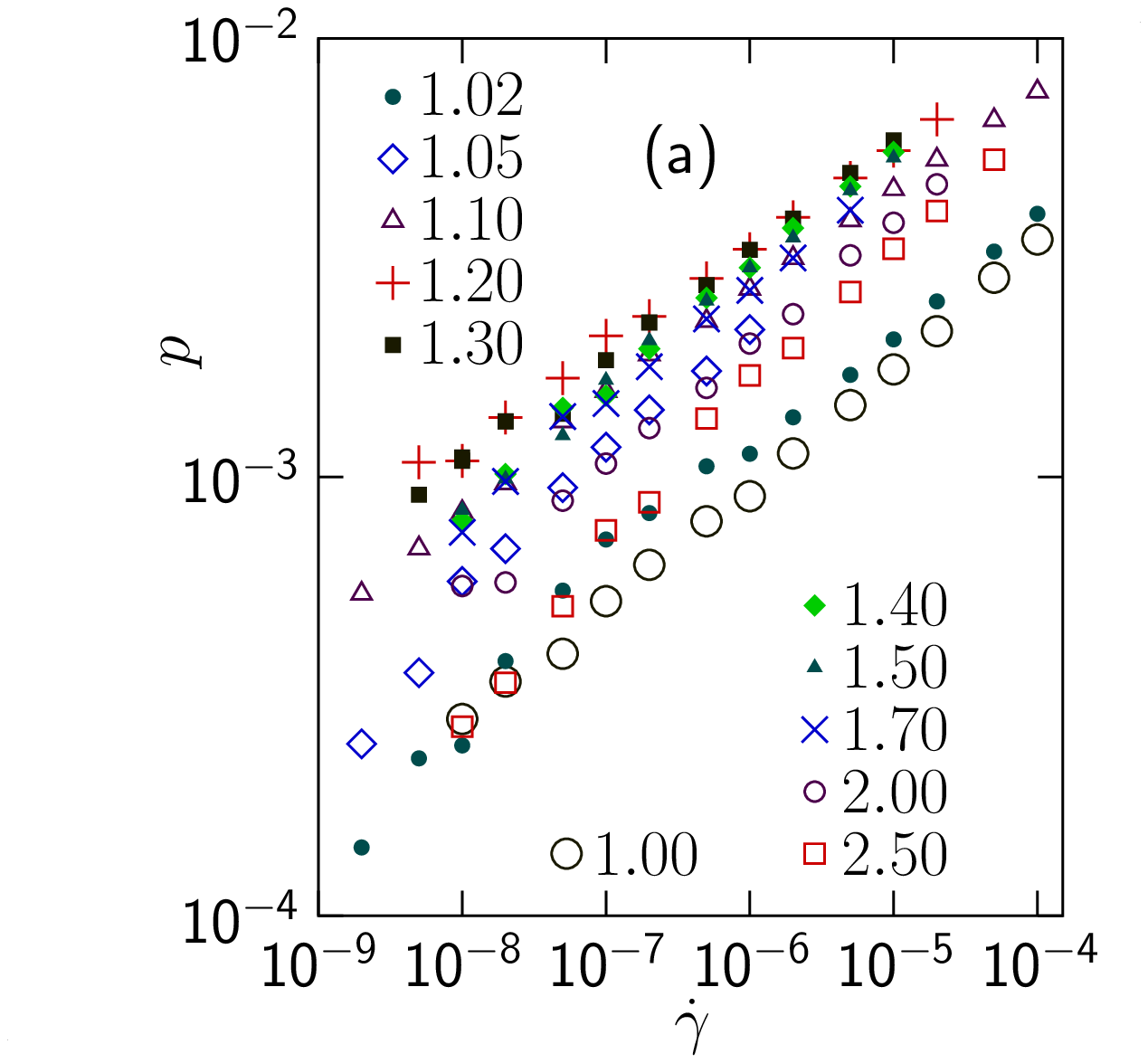}
  \includegraphics[bb=50 324 360 657, width=4.2cm]{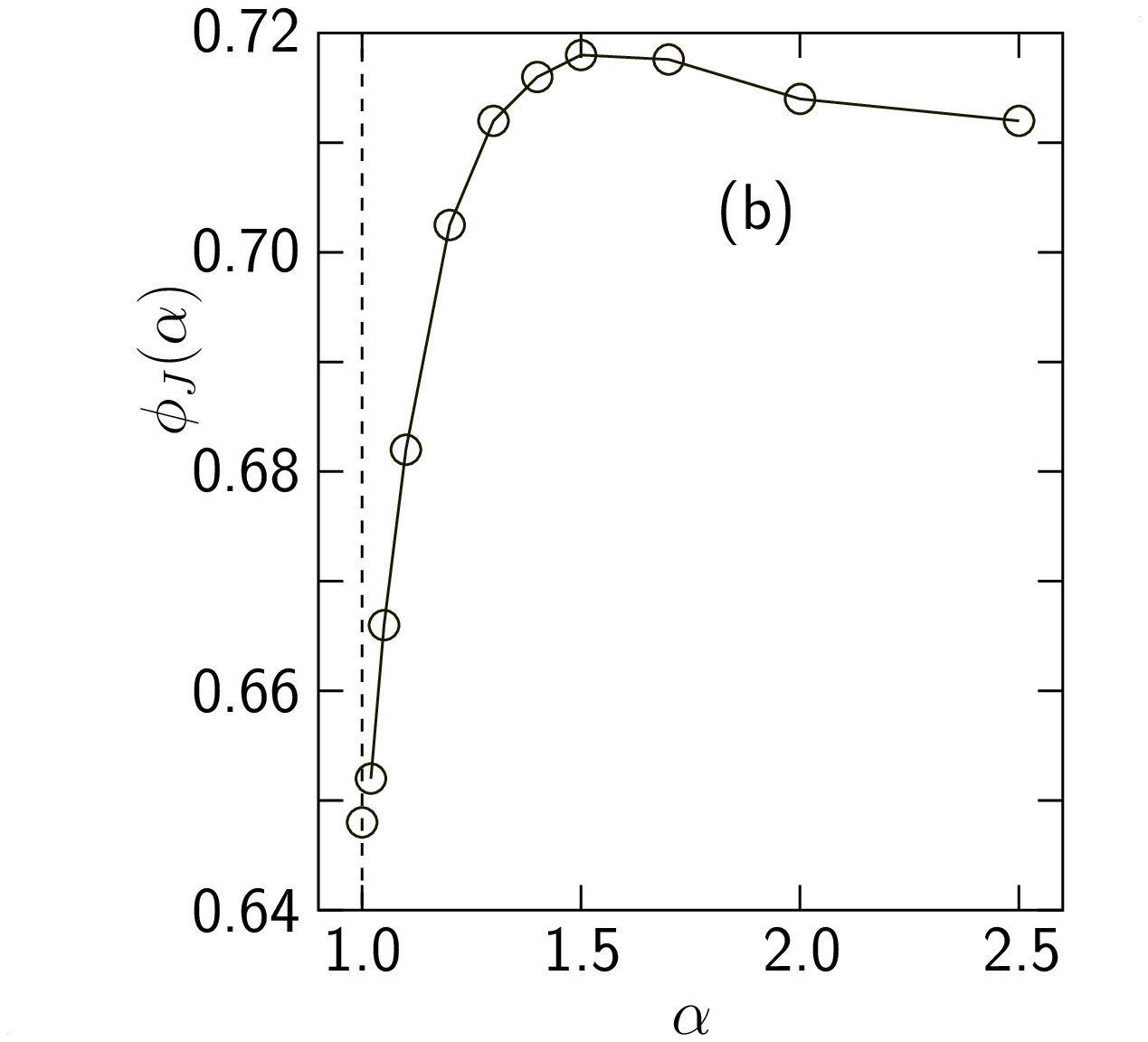}
  \caption{Determination of our approximate $\phi_J(\alpha)$. Panel (a) shows $p(\gdot)$
    for spheres (big open circles) and 10 different $\alpha>1$ at the respective
    $\phi_J(\alpha)$ chosen such that ecah data set is algebraic in $\gdot$ to a decent
    approximation. Panel (b) is $\phi_J(\alpha)$.}
  \label{fig:phistar}
\end{figure}

\Fig{phistar}(a) shows $p$ vs $\gdot$ both for spheres---big open circles---and for ten
different $\alpha>1$. We note that each $p(\gdot, \alpha)$ is algebraic and to a decent
approximation behaves the same as $p(\gdot, 1)$. Panel (b) is $\phi_J(\alpha)$.

From \Fig{phistar}(a) one can extract the dependence of $p$ on $\alpha$ at the respective
$\phi_J(\alpha)$ and constant $\gdot$ and it then appears that $p$ first increases, then
reaches a maximum at $\alpha=1.2$ or 1.3 and then eventually decreases as $\alpha$
increases further. This is thus in agreement with the conclusions of \Fig{vs-alpha}. It
should however be mentioned that this is only a tentative result since pressure is a
quantity that does depend strongly on $\phi$, which means that we cannot exclude the
possibility of a differently looking curve if the data had instead been taken at the true
$\phi_J(\alpha)$.

\subsection{Extrapolations to the $\gdot\to0$ limit}

\begin{figure}
  \includegraphics[bb=1 324 532 657, width=7cm]{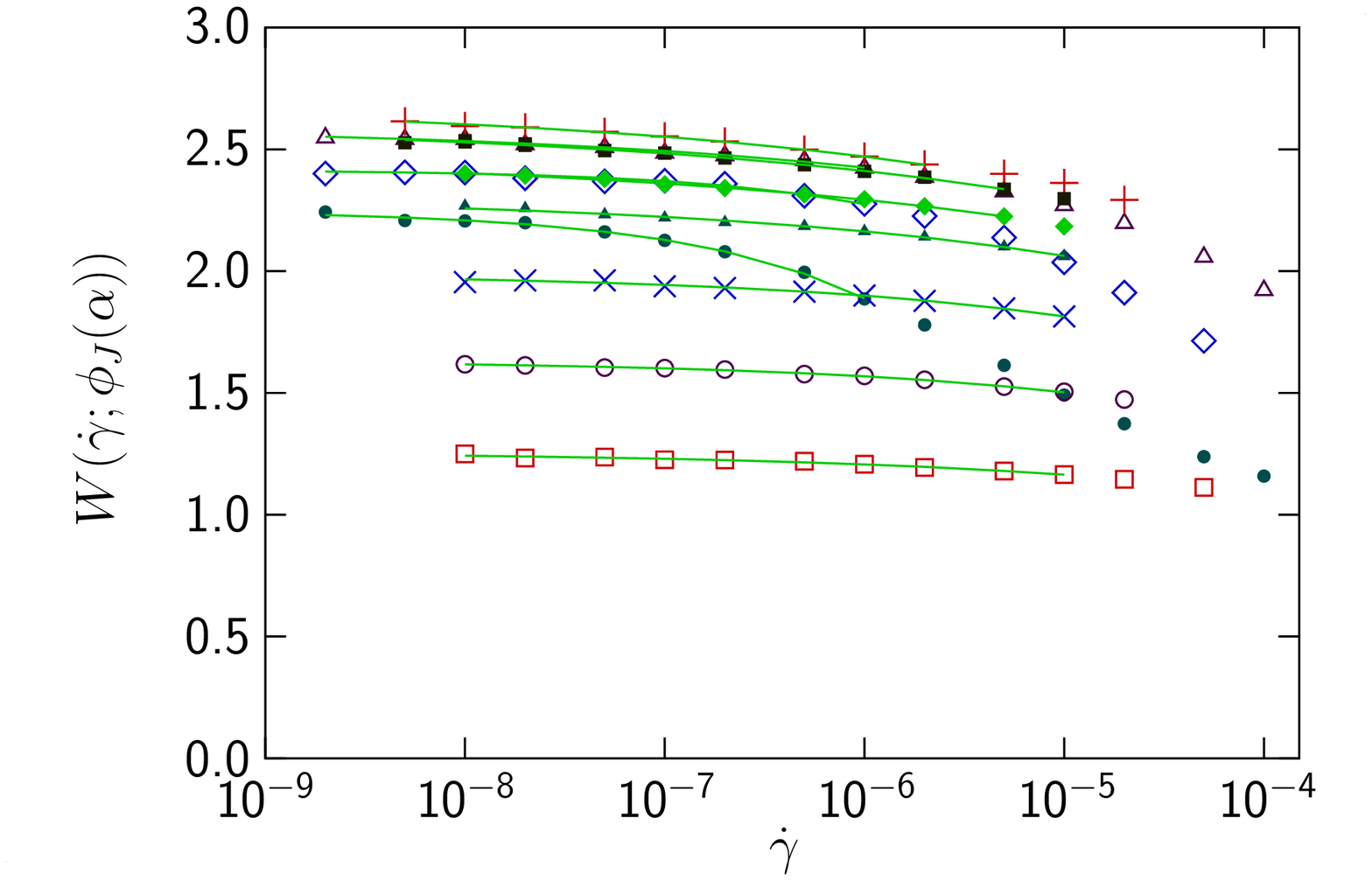}
  \includegraphics[bb=1 324 532 657, width=7cm]{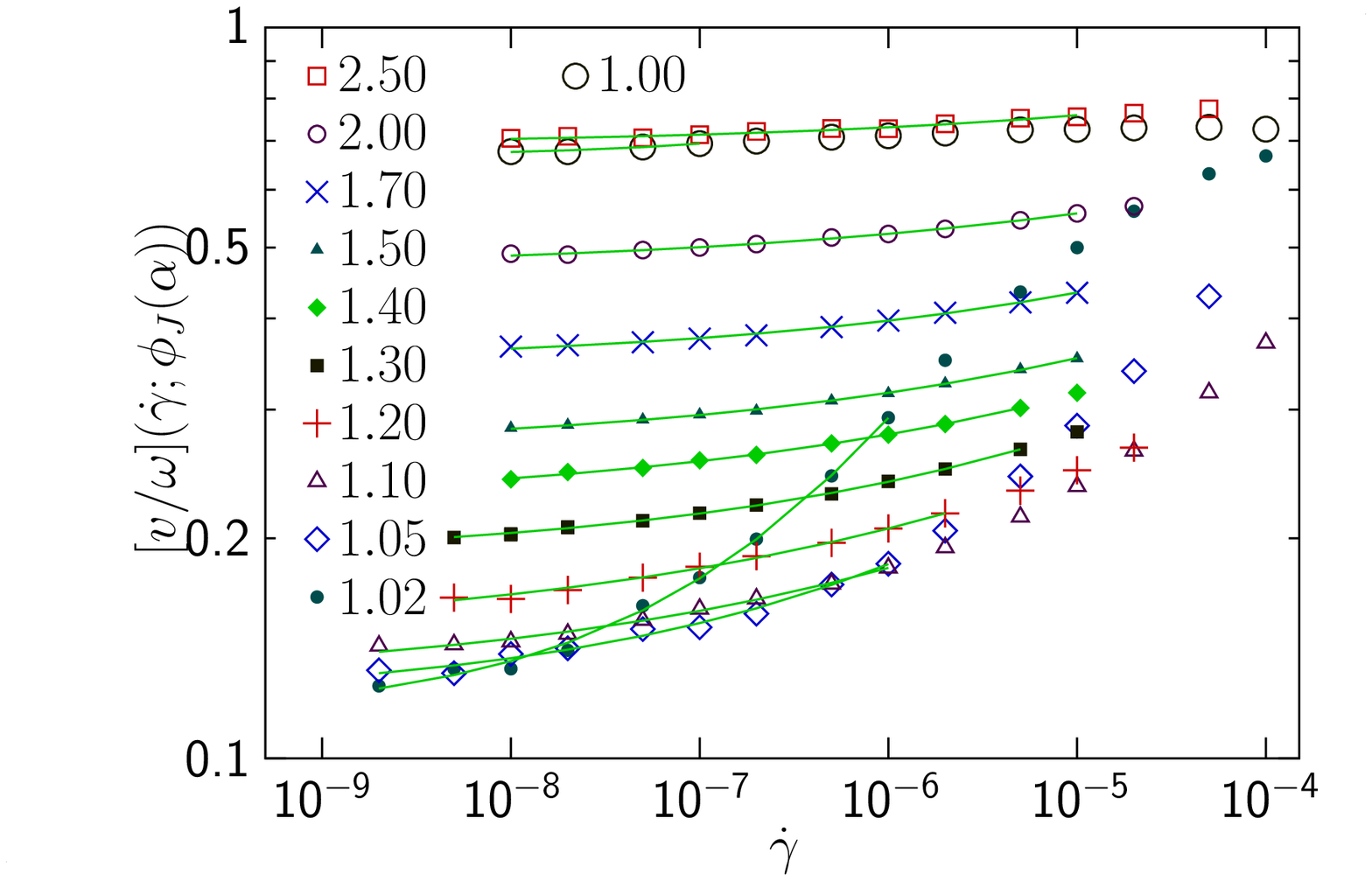}
  \includegraphics[bb=1 324 532 657, width=7cm]{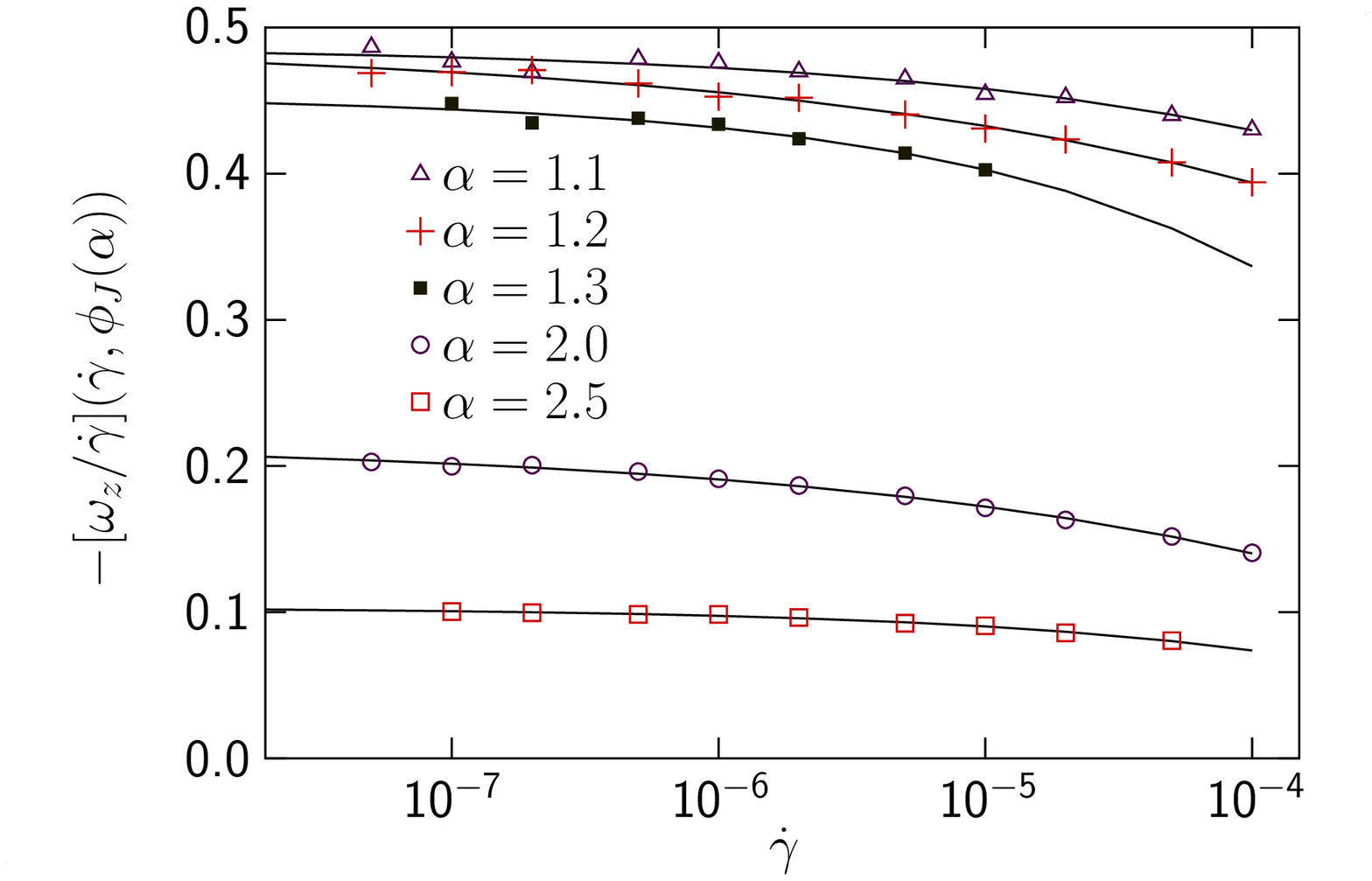}
  \caption{Extrapolations behind the data in \Fig{vs-alpha}(a), (b), and (d).}
  \label{fig:RandWandOmegaz}
\end{figure}

\Figure{RandWandOmegaz} shows extrapolations of three different quantities by fitting to
constant plus algebraic behavior. The corresponding constants, $W_0$, $[v/\omega]_0$ and
$[\omega/\gdot]_0$, are shown in \Fig{vs-alpha}(a), (b), and (d).

The constant shearing of the system in the $x$-$y$ plane leads to a rotational velocity
around the $z$ axis which for spheres is given by $\omega_z=-\gdot/2$. The determination
of $\omega_z$ is however difficult since the shear-driven rotation is a small signal
compared to the erratic particle rotations. To handle this we have here taken data from
some additional runs with $N=65536$ particles that give better statistics.  These runs
have however only been done for a few different $\alpha$, and that is the reason why there
are only five curves in \Fig{RandWandOmegaz}(c).

\bibliography{j}
\bibliographystyle{apsrev4-1}